\documentclass[%
 reprint,
 amsmath,amssymb,
 aps,
]{revtex4-2}
\DeclareUnicodeCharacter{0328}{\c{}}
\usepackage{adjustbox}
\usepackage{amsfonts}       
\usepackage{amsmath}
\usepackage{amssymb}
\usepackage{appendix}       
\usepackage{bm}             
\usepackage{caption}
\usepackage{dcolumn}        
\usepackage{float}
\usepackage{graphicx}       
\usepackage[hidelinks]{hyperref} 
\usepackage{cleveref}
\usepackage{mathtools}
\usepackage{paralist}
\usepackage{physics}
\usepackage{subfigure}
\usepackage{wrapfig}
\usepackage{xcolor}
\usepackage{xspace}
\usepackage{footnote}
\usepackage[justification=raggedright]{caption}
\usepackage{soul}
\usepackage{nameref}

\usepackage[utf8]{inputenc}
\usepackage[T1]{fontenc}
\usepackage{lmodern}

\begin{document}

\preprint{APS/123-QED}

\title{Running a six-qubit quantum circuit on a silicon spin qubit array}


\author{I. Fern\'andez de Fuentes\textsuperscript{1}}
\thanks{These authors contributed equally to this work.}
\author{E. Raymenants\textsuperscript{1}}
\thanks{These authors contributed equally to this work.}
\author{B. Undseth\textsuperscript{1}}
\author{O. Pietx-Casas\textsuperscript{1}}
\author{S. Philips\textsuperscript{1}}
\author{M. M\k{a}dzik\textsuperscript{1}}
\author{S. L. de Snoo\textsuperscript{1}}
\author{S. V. Amitonov\textsuperscript{2}}%
\author{L. Tryputen\textsuperscript{2}}%
\author{A. T. Schmitz\textsuperscript{3}}%
\author{A. Y. Matsuura\textsuperscript{3}}%
\author{G. Scappucci\textsuperscript{1}}%
\author{L. M. K. Vandersypen\textsuperscript{1}}%
\affiliation{\textsuperscript{1}QuTech and Kavli Institute of Nanoscience, Delft University of Technology, Lorentzweg 1, 2628 CJ Delft, The Netherlands
}
\affiliation{\textsuperscript{2}
 QuTech and Netherlands Organization for Applied Scientific Research (TNO), Delft, The Netherlands
}
\affiliation{\textsuperscript{3}
 Intel Corporation, Hillsboro, OR, United States
}

\date{\today}

\begin{abstract}
The simplicity of encoding a qubit in the state of a single electron spin and the potential for their integration into industry-standard microchips continue to drive the field of semiconductor-based quantum computing.  
However, after decades of progress, validating universal logic in these platforms has advanced little beyond first-principles demonstrations of meeting the DiVincenzo criteria. Case in point, and specifically for silicon-based quantum dots, three-qubit algorithms have been the upper limit to date, despite the availability of devices containing more  qubits. In this work, we fully exploit the capacity of a spin-qubit array and implement a six qubit quantum circuit, the largest utilizing semiconductor quantum technology. By programming the quantum processor, we execute quantum circuits across all permutations of three, four, five, and six neighbouring qubits, demonstrating successful programmable multi-qubit operation throughout the array. The results reveal that, despite the high quality of individual units, errors quickly accumulate when combining all of them in a quantum circuit. This work highlights the necessity to minimize idling times through simultaneous operations, boost dephasing times, and consistently improve state preparation and measurement fidelities.
\end{abstract}

\maketitle

\section{Introduction}
\label{sec:Introduction}

The prospect of achieving quantum advantage and real-world applications drives the rapidly evolving field of quantum technologies. For quantum computing, a major challenge in outpacing classical devices is scaling the number of qubits while maintaining high-fidelity initialization, control, and readout of the qubits. 
 
Over the past decades, platforms based on superconducting qubits\,\cite{acharya2024quantum}, neutral atoms\,
\cite{bluvstein2024logical} and trapped ions~\cite{decross2024computational,paetznick2024demonstration} have come forward increasing the number and quality of qubits, enabling important milestones such as the first demonstrations of logical encoding for quantum error correction\,\cite{krinner2022realizing,acharya2024quantum,paetznick2024demonstration} or experimental studies of many-body phenomena\,\cite{ebadi2021quantum,bluvstein2021controlling,andersen2025thermalization}.

Following in their steps, quantum hardware based on semiconductor spin qubits is raising the stakes by leveraging industry-standard fabrication techniques to scale up\,\cite{gonzalez2021scaling,zwerver2022qubits,neyens2024probing,elsayed2024low,george202412,ha2025two}. 
Considerable focus has been given to validate these platforms by demonstrating and refining high-fidelity single-\,\cite{steinacker2024300} and two-qubit gates\,\cite{xue2022quantum,noiri2022fast,tanttu2024assessment}, as well as initialization and readout\,\cite{huang2024high,takeda2024rapid,steinacker2024300}. This progress has enabled the execution of realistic quantum circuits with two\,\cite{watson2018programmable}, three~\cite{takeda2021quantum,van2022phase,thorvaldson2025grover} and four qubits\,\cite{hendrickx2021four,zhang2024universal}. In an array of ten quantum dots,  single-qubit operations were characterized one at a time~\cite{john2024two}. Finally, in a six-qubit device, initialization, single- and two-qubit gates, and readout were simultaneously tuned up across the full system, although entangling circuits were implemented with up to three qubits only~\cite{philips2022universal}.

In this work, we go a step further and assess the performance of a six-qubit array when all the available units in the quantum processor are involved in the quantum circuit. This is a crucial aspect, as a potential bottleneck for the scalability of these noisy intermediate-scale devices arises precisely from this collective operation: as more units are added, errors accumulate, overall processor operation slows down, and the available coherence budget for quantum operations becomes limiting.

We utilize six single-spin qubits arranged in a linear array of quantum dots to run a multiqubit quantum circuit inspired by a protocol designed to investigate dynamical quantum phase transitions in Ising-like systems. Leveraging single-qubit control through addressable microwave driving and tunable two-qubit exchange interactions, we showcase the quantum processor’s programmability in operating with any combination of three, four, five or six qubits. To demonstrate and validate this capability, we employ an protocol that can flexibly accommodate an increasing number of qubits with a well-understood expected scaling of its output. This enables us to identify key bottlenecks and limitations in the current generation of these platforms, particularly with regards to runtime and state preparation and measurement (SPAM) errors, while also outlining potential pathways for improvement.

This manuscript is organized as follows. In Section \,\ref{sec:hardware} we present details on the quantum processor platform and its operation. The implemented quantum circuit is described in Section \,\ref{sec:algorithm}. In Section \,\ref{sec:loschmidt_magnetization} we report the experimental results obtained from running the quantum circuit and discuss the sources of error which limit the performance of the quantum processor. We provide further benchmarks through Quantum State Tomography in Sec.\,\ref{sec:state_tomo}. Finally, we discuss the implications of the results in Section \,\ref{sec:Discussion}.

\section{\label{sec:hardware} Device, initialization and readout}

\begin{figure*}[t]
  \centering
  \includegraphics[width=\textwidth]{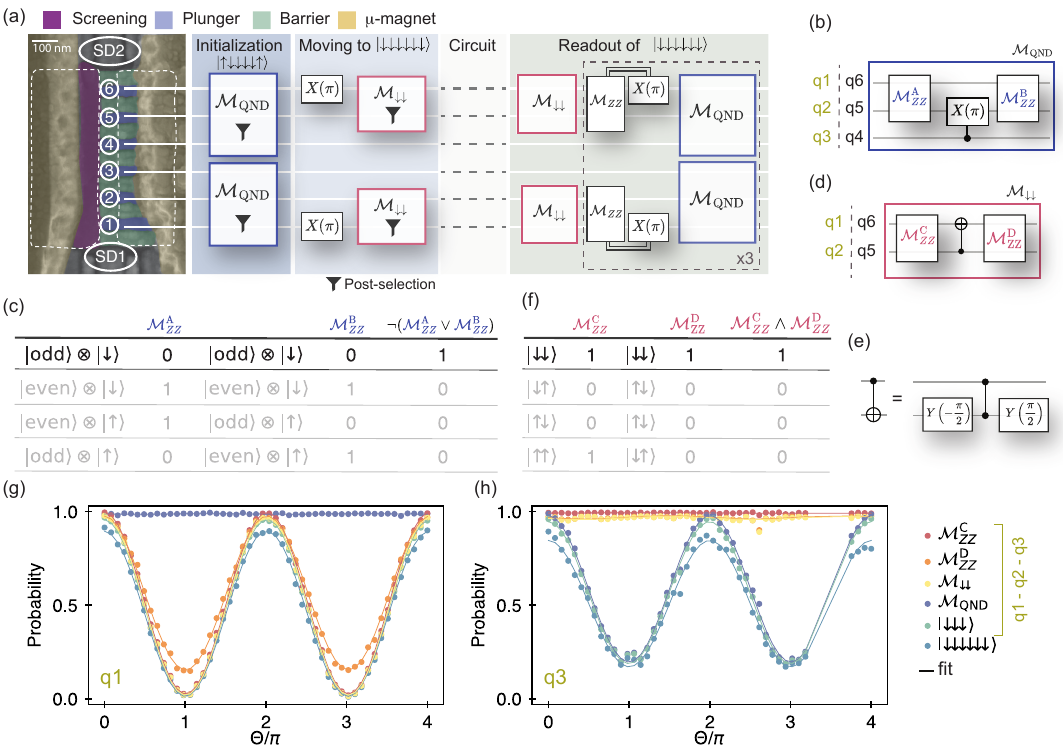}
  \caption{\textbf{Device initialization and readout scheme.} \textbf{(a)} False-colored SEM image of a device nominally identical to the one used, alongside the quantum circuit schematic for initializing and reading out all six qubits. Electrons are confined in quantum dots centered beneath the plunger gates and in between two screening gates (white dashed contours). The interdot tunnel coupling is controlled via the barrier gates. The cobalt micromagnet deposited on top creates a magnetic field gradient enabling single-qubit addressability while also producing synthetic spin-orbit coupling for single-spin control via electric-dipole spin resonance (EDSR) by driving a screening gate. \textbf{(b)} Quantum circuit used to construct the measurement operator $\mathcal{M}_{\rm{QND}}$ 
     by means of a quantum non-demolition measurement. \textbf{(c)} The corresponding truth table based on the parity measurements $\mathcal{M}_{ZZ}^{\rm{A}}$ and $\mathcal{M}_{ZZ}^{\rm{B}}$. Highlighted is the state that we post-select after measurement for initialization based on the operator $\mathcal{M}_{Z}^{\rm{QND}}$ which is extracted as shown in the last column. \textbf{(d)} Quantum circuit to build the operator $\mathcal{M}_{\downarrow\downarrow}$. Here, the parity measurements $\mathcal{M}_{ZZ}^{\rm{C}}$ and $\mathcal{M}_{ZZ}^{\rm{D}}$ allow discrimination of the $\ket{\downarrow\downarrow}$ state using the truth table in \textbf{(f)}, corresponding to two even outcomes. \textbf{(e)} Compilation of the CNOT gate used in \textbf{(d)}.  \textbf{(f)} Truth table for the parity readouts $\mathcal{M}_{ZZ}^{\rm{C}}$ and $\mathcal{M}_{ZZ}^{\rm{D}}$. The notation here assumes that the arrows in the ket correspond to qubits 1 or 6 for the first arrow and qubits 2 or 5 for the second. Note that the state $\ket{\downarrow\uparrow}$ decays to the $\ket{\uparrow\downarrow}$ after the first PSB.  Highlighted is the state that we post-select after measurement for initialization.  The operator $\mathcal{M}_{ZZ}^{\downarrow\downarrow}$ is extracted through the logical combination shown in the last column. \textbf{(g-h)} Rabi experiments obtained for qubits 1 and 3 after preparing the array in the $\ket{\downarrow}^{\otimes 6}$ state and applying a resonant drive to achieve a calibrated rotation of an angle $\theta$ of \textbf{(g)} qubit 1 and \textbf{(h)} qubit 3. The various data points (see legend) show all intermediate observables as well as the probability for obtaining $\ket{\downarrow}^{\otimes 3}$ in the subspace 1-2-3, and the return probability to the $\ket{\downarrow}^{\otimes 6}$ state. The solid lines correspond to fits using sinusoidal functions. }
  \label{fig:Figure1}
\end{figure*}

The six electron spin qubits are hosted in an array of six quantum dots that are electrostatically defined within a $^{28}$Si/SiGe heterostructure. The dots are formed below the plunger gates of the device depicted in Fig.\,\ref{fig:Figure1}\,a. Details on the design, operation and calibration of the system are extensively covered in\,\cite{philips2022universal}. However, we revisit and extend the readout and initialization schemes as they must be adapted for the experiments here. Additionally, we intentionally operate the device at 200 mK to reduce unwanted heating effects induced by the delivery of microwave signals during qubit control\,\cite{undseth2023hotter}.

The initialization and readout sequences are depicted in the blue and grey shaded areas in Fig.\,\ref{fig:Figure1}\,a, respectively. For compactness, the steps for qubits 1-3 and for qubits 4-6 are illustrated as occurring simultaneously in the diagram, but in practice they are executed sequentially. 

Readout relies on two charge sensing dots, one on each end of the array and labeled as `SD' in the false-colored SEM image. Each sensing dot is used for reading out the spin state of the two nearest confined electrons, namely pair 1-2 and pair 5-6, through parity-mode Pauli Spin Blockade (PSB)\,\cite{seedhouse2021pauli} (see Methods). Note that throughout the text, we may occasionally refer to `parity-mode PSB' as simply `PSB'. The associated measurement operator $\mathcal{M}_{ZZ} = (1+ZZ)/2$, returns  a 0 or a 1 depending on whether the spins are in an antiparallel (odd) or parallel (even) configuration, respectively. Due to the specifics of the PSB readout used here, the post-measurement state for an odd parity outcome will be the $\ket{\uparrow\downarrow}$ ($\ket{\downarrow\uparrow}$) state with high probability for qubits 1-2 (5-6), regardless of the pre-measurement state (see Methods). For the even parity outcomes, the post-measurement state with high probability is the same as the pre-measurement state.

As charge sensing is less efficient for the inner dots due to their greater distance to the sensing dots, the spin state of qubit 3\,(4) is read out via a Quantum Non-Demolition (QND) measurement\,\cite{nakajima2019quantum} whereby the state of qubit 3\,(4) is mapped onto that of qubit 2\,(5) by a conditional $X(\pi)$-rotation (CROT), which can be subsequently measured via PSB as shown in Fig.\,\ref{fig:Figure1}\,b. With a reference parity measurement, we can use the probability of parity flipping as a proxy for the  measurement operator $\mathcal{M}_Z = (1+Z)/2$ on qubits 3 and 4, as illustrated in the truth table in Fig.\,\ref{fig:Figure1}\,c. Furthermore, the QND measurement procedure can be repeated to boost the measurement fidelity by taking a majority vote of repeated measurements.

Applying the parity measurement protocol to qubits 1-2 and 5-6 and the QND measurement protocol to qubits 3 and 4 within a single measurement cycle yields 4 bits of information regarding the state of the full array provided by the measurement operators $\mathcal{M}_{Z_1Z_2}\;,\mathcal{M}_{Z_3},\; \mathcal{M}_{Z_4}\; \rm{and}\; \mathcal{M}_{Z_5Z_6}$~\cite{philips2022universal}. 

For the successful implementation of the quantum circuit, as will be detailed in Section \ref{sec:algorithm}, an extended readout scheme is necessary to obtain the so-called \textit{return probability}. This probability is defined as the overlap of the initial and final quantum states, $|\bra{\psi_{0}}\ket{\psi_{\rm f}}|^2$. Specifically, we aim at initializing and measuring the return probability of the $\ket{\psi_{0}} = \ket{\downarrow}^{\otimes N}$ state, with $N$ the number of qubits. However, in the current PSB readout scheme for qubit pairs 1-2 and 5-6~\cite{philips2022universal}, it is inherently not possible to discriminate between $\ket{\downarrow\downarrow}$ and $\ket{\uparrow\uparrow}$, and also not between $\ket{\uparrow\downarrow}$ and $\ket{\downarrow\uparrow}$. 
As a result, the return probability to the  $\ket{\downarrow}^{\otimes N}$ state cannot be directly measured. 

To overcome this limitation, two-qubit logic can be employed to isolate the $\ket{\downarrow}^{\otimes N}$ state from all the other states. 
This approach is illustrated in the circuit schematics from Fig.\,\ref{fig:Figure1}\,d. The scheme begins with a first measurement $\mathcal{M}_{ZZ}^{\rm{C}}$, which determines if the spin parity of qubit pair 1-2 is even or odd (the scheme for qubit pair 5-6 is analogous). Next, a controlled-NOT (CNOT) gate is applied between the parity pairs using qubit 2 as the control qubit and qubit 1 as the target qubit. 

We compile the CNOT using a sequence of operations which leverages the native two-qubit $ZZ$ interaction as shown schematically in Fig.\,\ref{fig:Figure1}\,e (see Methods).
A second parity measurement $\mathcal{M}_{ZZ}^{\rm{D}}$ follows the CNOT operation, leading to the possible outcomes listed in Fig.\,\ref{fig:Figure1}\,f. By selecting outcomes where the parity is even in both $\mathcal{M}_{ZZ}^{\rm{C}}$ and  $\mathcal{M}_{ZZ}^{\rm{D}}$, the state $\ket{\downarrow\downarrow}$ can be isolated resulting in an effective measurement of the observable $\mathcal{M}_{\downarrow\downarrow} = \ket{\downarrow\downarrow}\bra{\downarrow\downarrow}$ for the first two and the last two qubits. 
As before, measurement of qubits 3 and 4 is implemented directly through $\mathcal{M}_{\rm QND}$. While not strictly necessary, we incorporate an additional step within the readout sequence, marked by the rotation of qubits 1 and 6 conditional on an even parity outcome, which takes the pairs to the lower-energy antiparallel configuration before $\mathcal{M}_{\rm QND}$ readout. This approach is motivated by the fact that it yields a higher number of retained traces after the initialization sequence of the next round. 

We prepare the qubits using initialization by measurement and post-selection, making use of similar sequences as during readout (light and dark blue region in Fig.\,\ref{fig:Figure1}\,a). First, we perform a QND measurement on qubits 3 and 4 and post-select for $\ket{\downarrow}$ outcomes for these qubits and for odd outcomes on qubits 1-2 and 5-6 during $\mathcal{M}_{ZZ}^{\rm{B}}$ readout. This results in the post-selected post-measurement state $\ket{\uparrow\downarrow\downarrow\downarrow\downarrow\uparrow}$~\cite{philips2022universal}. After flipping qubits 1 and 6 with a resonant $\pi$-pulse, the $\ket{\downarrow\downarrow\downarrow\downarrow\downarrow\downarrow}$ state is prepared. While not strictly necessary, we add another set of parity measurements to confirm the outer qubit pairs are set to $\ket{\downarrow\downarrow}$. Unless stated otherwise, we prepare and measure all six qubits in all the experiments, even when not all qubits are involved in the circuit, to ensure that state preparation and measurement (SPAM) errors are similar. Additionally, systematic sensor drifts from residual microwave heating are mitigated during postprocessing (Supplementary Material S1).

To validate the extended initialization and readout scheme, we resonantly drive Rabi oscillations on individual qubits using electric-dipole spin resonance and inspect the probability of measuring individual and composite observables. The results for driving qubits 1 or 3 are depicted in Fig.\,\ref{fig:Figure1}\,g-h as an example, and the same experiments for all six qubits can be found in Supplementary Material S2. 
 Looking at the experimental results in Fig.\,\ref{fig:Figure1}\,g, we observe oscillations in the $\mathcal{M}_{\downarrow\downarrow}$ and $\ket{\downarrow}^{\otimes 3}$ probabilities while the $\mathcal{M}_{\rm{QND}}$ probability stays constant, which is consistent with coherently driving qubit 1 only. Analogously, the data in Fig\,\ref{fig:Figure1}\,h shows that $\mathcal{M}_{ZZ}^{\rm{C}}$, $\mathcal{M}_{ZZ}^{\rm{D}}$ and consequently $\mathcal{M}_{\downarrow\downarrow}$ remain flat, while $\mathcal{M}_{\rm{QND}}$  oscillates, verifying that the oscillations in the $\ket{\downarrow}^{\otimes 3}$ probability are purely due to changes in the state of qubit 3. In both cases, also the $\ket{\downarrow}^{\otimes 6}$ probability oscillates, as expected. In interpreting the data, it is good to keep in mind that the AND logic between the outcomes of $\mathcal{M}_{ZZ}^{\rm{C}}$ and $\mathcal{M}_{ZZ}^{\rm{D}}$ inherently biases the results towards lower values in the presence of SPAM errors (see Supplementary Material S3 for a rigorous derivation). To avoid overcrowding the figures, we do not show the intermediate readouts on the other half of the array in Figs.\,\ref{fig:Figure1}\,g-h. Naturally, a reduction in the oscillation amplitude for the $\ket{\downarrow}^{\otimes 6}$ is expected when combining the readout measurements from qubits 4,\,5 and\,6 from accumulated SPAM errors decreasing the visibility by approximately 6\% for qubit 1 and 10\% for qubit 3.  

\begin{figure*}[t!]
  \includegraphics[width=0.9\textwidth]{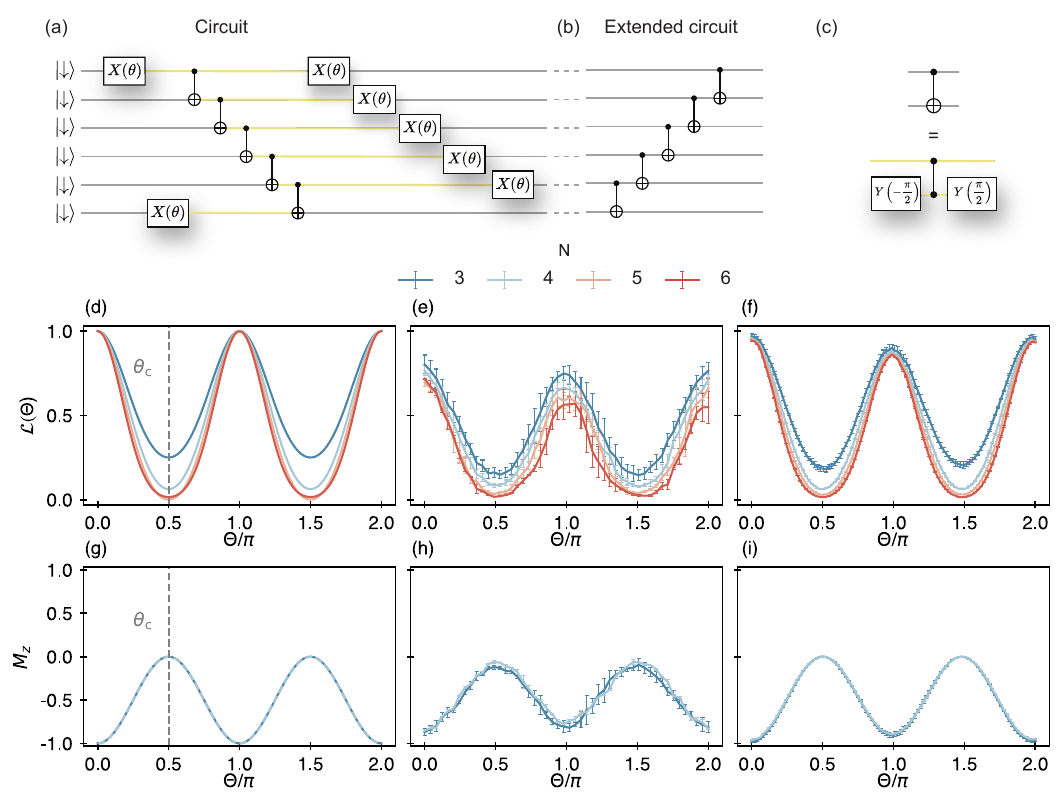}
  \caption{\textbf{Quantum circuit diagram and results} \textbf{(a)}  Quantum circuit diagram to perform a quantum quench. This circuit is to be executed in between initialization of the system in $\ket{\downarrow\downarrow\downarrow\downarrow\downarrow\downarrow}$ and the measurement of the return probability, shown in Fig.~\ref{fig:Figure1}\textbf{(a)}. The yellow coloring highlights idle periods of the qubits between gate operations, also shown for \textbf{(c)}. \textbf{(b)} Quantum circuit extension needed for the initial-state-independent version of the circuit. \textbf{(c)} Quantum circuit showing how the CNOT operations in \textbf{(a)} are compiled. \textbf{(d-f)} Return probability $ \cal L$ as function of $ \theta $ obtained from \textbf{(d)} a numerical simulation not accounting for noise, \textbf{(e)} experiments, for each $N$ averaged over all possible permutations of neighboring qubits, and \textbf{(f)} numerical simulations including dephasing errors from quasi-static noise. (\textbf{g-i}) Magnetization defined as $M_z = \frac{1}{N} \sum_i^N  Z_i $ for $N$ is 3 or 4, obtained from \textbf{(g)} a numerical simulation not accounting for noise, \textbf{(h)} experiments, for each $N$ averaged over all possible permutations of neighbouring qubits among qubits 2, 3, 4 and 5, and \textbf{(i)} numerical simulations including dephasing errors from quasi-static noise. The error bars in the experimental panels correspond to one standard deviation.}
  \label{fig:Figure2}
\end{figure*}

\section{\label{sec:algorithm} Quantum Circuit Description} 
Having established the initialization and readout procedure, we now turn to detailing the quantum circuit used in this work. The circuit in Fig.\,\ref{fig:Figure2}\,a depicts a digital protocol where a quantum state described by the ground state of an initial Hamiltonian $H_0 =h\sum_{i =1}^{N}Z_i$ undergoes a quantum quench by suddenly activating the interaction Hamiltonian $H_{\rm quench}(h = 0) = J_0\left[\sum_{i=1}^{N-1}X_{i}X_{i+1} + X_1+X_N\right]$. Here, $h$ denotes the strength of the on-site longitudinal field, while $J_0$ governs both the strength of the nearest-neighbor coupling and the boundary conditions via the edge terms  $X_{1}$ and $X_N$ (see Supplemetary Material S4\,A). The corresponding out-of-equilibrium dynamics can be effectively mapped to a circuit composed of single- and two-qubit gates as displayed, without the need of Trotterization. 
Note that we do not need to implement the second set of CNOT gates in Fig.\,\ref{fig:Figure2}\,b because the experiments are focused on the case where the initial state $\ket{\psi_0} = \ket{\downarrow}^{\otimes N}$, the ground state of $H_0$. In this scenario, the evolution under the $XX$ interactions can be simplified to the circuit in Fig.\,\ref{fig:Figure2}\,a. A detailed derivation for the mapping of the quenching Hamiltonian and the aforementioned simplification can be found in Supplementary Material S4\,B. Our motivation to choose this particular protocol stems from its simplicity while still incorporating both single-qubit operations on all qubits and two-qubit interactions between all neighboring qubits in a linear chain. Furthermore, in the course of the circuit execution, all qubits are expected to become entangled. This scheme thus constitutes a meaningful first test of the full six-qubit device under a realistic workload. 

Fig.\,\ref{fig:Figure2}\,a illustrates the circuit when all six qubits in the array are used. The adaptation of the quantum circuit for $N = \{3,4,5\}$ is achieved by omitting one, two or three of the inner circuit channels, respectively.
Time evolution is parameterized by the angle $\theta$, which we can experimentally vary by sweeping the duration $t_{\rm MW}$ of a microwave burst  on resonance with the target qubit, such that $\theta = \pi\,t_{\rm MW}/t_{\rm \pi} $ with $t_{\rm \pi}$ the time needed to perform a full flip  of the spin, ranging from 94-250 ns. Entanglement is distributed across the array via the CNOT gates, compiled as shown in the schematic of Fig.\,\ref{fig:Figure2}\,c with a typical duration of 100-800 ns.

As introduced in Section \,\ref{sec:hardware}, in these experiments we are interested in tracking the overlap of the initial state $\ket{\psi_0}$ with a final state—the time-evolved state $\ket{\psi(t)}$ in this case—referred to as the return probability. This metric, also known as the Loschmidt echo amplitude, is commonly used to probe the complex dynamics of interacting quantum systems\,\cite{jurcevic2017observation,xu2020probing}, and it is formally defined as $\mathcal{L}(\theta) = |\bra{\psi_0}\ket{\psi(\theta)}|^2=|\bra{\psi_0}e^{-iHt}\ket{\psi_0}|^2$. The chosen protocol compares with models which have been previously studied in the context of dynamical quantum phase transitions (DQPTs)\,\cite{jurcevic2017observation,kim2023scalable}. While our aim is not to focus on the physical implications within the framework of quantum phase transitions, we want to draw attention to the scaling of the Loschmidt echo amplitude when the number of qubits, N, involved in the dynamics change, which can be expressed analytically as (see Supplementary Section S4\,C):
\begin{align}
\mathcal{L}(\theta) =  \left|\cos\left(\frac{\theta}{2}\right)^{N+1}+i^{N+1} \sin\left(\frac{\theta}{2}\right)^{N+1}\right|^{2}.  
\label{eq:lochsmidt_analytic}
\end{align}
From Eq.\,\ref{eq:lochsmidt_analytic}, we observe that the Loschmidt echo amplitude exhibits a periodicity of  $\theta = \pi$ and depends on $N$. Fig.\,\ref{fig:Figure2}\,d illustrates the expected $\mathcal{L}(\theta)$ for  $N$ varying from three to six. As $N$ increases, the Loschmidt echo exhibits a broader and deeper minimum, indicating that $\ket{\psi(\theta)}$ evolves faster and farther away from the initial $\ket{\psi_0}$. In all cases, $\mathcal{L}(\theta)$ reaches a minimum at $\theta = \pi/2$, referred to as the \textit{critical time} $\theta_{\rm c}$ in the DQPT terminology. An analogy to equilibrium statistical physics is often drawn for these critical points, particularly relating them to phase transitions in which thermodynamical properties become non-analytical (see Supplementary Material S4\,C for details). Lastly, we want to highlight that the periodicity observed in Fig.\,\ref{fig:Figure2}\,d differs from what one would expect from a periodicity of $2\pi$ based solely on single-qubit rotations, $X(\theta)$. The two-qubit interactions are responsible for the periodicity of $\theta = \pi$ (See Supplementary S5 for more details).

\section{Experimental results}
\subsection{\label{sec:loschmidt_magnetization} Measurements of return probability and magnetization}

Using the methodology for state preparation and measurement described in Section\,\ref{sec:hardware}, we run the quantum circuit presented in Fig.\,\ref{fig:Figure2}\,a  for different system sizes, $N \in \{3,4,5 ,6\}$. We note that the operations are performed sequentially in the current implementation of the circuit, as illustrated in Fig.\,\ref{fig:Figure2}\,a by appending the gates in the circuit. The experimental results are presented in Fig.\,\ref{fig:Figure2}\,e, featuring the Loschmidt echo amplitude as a function of  angle $\theta$ and number of particles $N$. Importantly, these datasets were collected on the same day after a full calibration round as described in the Methods Section \ref{sec:calibrations}.

Each trace in Fig.\,\ref{fig:Figure2}\,e displays the average result after executing the circuit across all possible qubit combinations for varying system sizes. For instance, for $N=5$, we average over four datasets, involving either qubits 1-5 or qubits 2-6, and with the five retained qubits ordered sequentially or in reverse. Acknowledging that individual qubit combinations may exhibit varying performance, we use this approach as an estimation of the \textit{average behavior} of the quantum processor. The separate measurement results for all cases are presented in Fig.\,S5 of the Supplementary Material. There, we also include the raw data for which the readout signal was not re-thresholded to mitigate sensor drifts due to heating. The results feature the oscillations as a function of angle $\theta$, as predicted by Eq.\,\ref{eq:lochsmidt_analytic}, showing that the quantum state evolves periodically through regions of maximum and minimum overlap with the initial state with period $\theta = \pi $ and critical angle $\theta_{\rm c}=\frac{\pi}{2}$. As previously discussed, this indicates that interactions across the entirety of the array can indeed be controllably activated to realize succesive two-qubit gates. We can also appreciate that increasing $N$ results in broader dips, consistent with predictions.

\begin{figure*}[t]
\includegraphics[width=\textwidth]{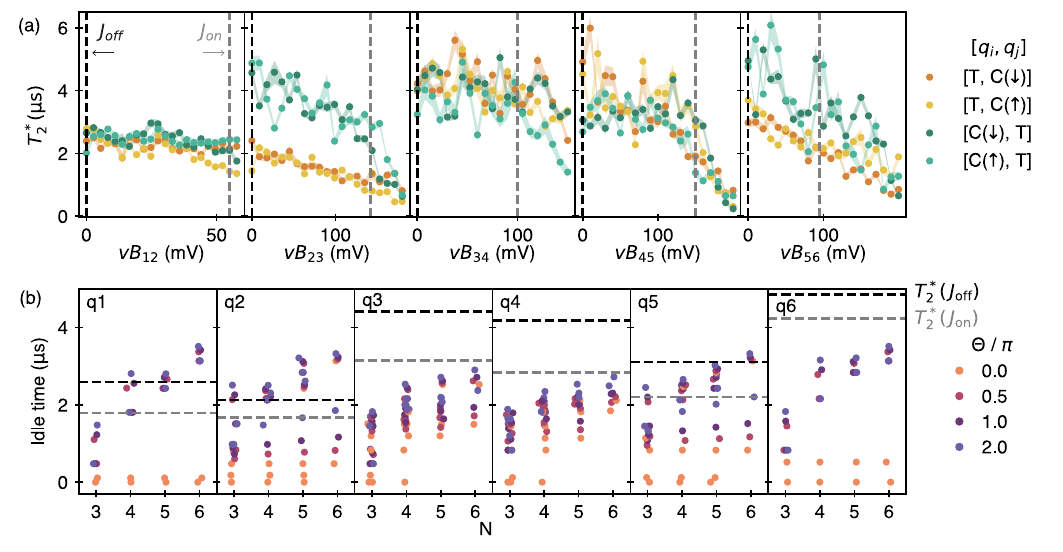}
  \caption{\textbf{Idling timescales and dephasing times} \textbf{(a)} $T_{2}^{*}$ times extracted from Ramsey experiments for all six qubits, conditional on the state ($\uparrow$ or $\downarrow$) of a neighbouring qubit, and  as a function of virtual barrier pulse amplitude $vB_{ij}$. Each panel shows the $T_2^*$ values for the two qubits $i$ and $j$ adjacent to the virtual barrier gate $vB_{ij}$. The labels T and C refer to the target and control qubit, respectively.  The vertical dashed lines mark the values of $vB_{\rm{ij}}$ at which single- ($J_{\rm{off}}$) and two-qubit ($J_{\rm{on}}$) operations are executed. \textbf{(b)} Total idle times experienced by individual qubits 1-6 throughout the circuit as a function of the number of qubits involved in the circuit $N$. For clarity, a modest amount of jitter has been added to the $x$-axis to separate the data points. As expected, idle times increase with $N$, as quantum logic is applied sequentially. Fewer data points are available for the edge qubits, due to a reduced number of possible combinations. For instance, qubit 3 can be involved in running the circuit for $N=3$ using qubits 1, 2, 3; 3, 2, 1; 2, 3, 4; 4, 3, 2; 3, 4, 5 and 5, 4, 3; while qubit 1 can only be involved in circuits using qubits 1, 2, 3 and 3, 2, 1. The $T_2^{*}$ values for the conditions $J_{\rm{off}}$ and $J_{\rm{on}}$ in \textbf{(a)} (averaged over the four sets of datapoints) are included as a visual guide to compare the decoherence times against the idling times, showing that the idle times are in many cases comparable to or even surpass $T_2^{*}$, in particular for qubits 1, 2 and 5.}
  \label{fig:Figure3}
\end{figure*}

Despite the overall similarities between theory and experiment, it is noticeable that the experiments feature an $N$-dependent offset of the curves at $\theta = 0$ and overall loss of visibility. In addition, we observe a decay increasing with $\theta$ especially for larger $N$. The $N$-dependent offset is not likely a consequence of state preparation and measurement errors (SPAM) accumulating as $N$ increases, as all qubits are prepared and read out independent of the number of qubits that are used in the circuit. However, we expect that SPAM contributes to an overall reduction in the amplitude of the oscillations when benchmarked against the theoretical estimations in Fig.\,\ref{fig:Figure2}\,d, an effect that was already observed in the Rabi experiments of Fig.\,\ref{fig:Figure1}\,g-h.

Looking at the impact of gate imperfections, at $\theta=0$, we naively may expect that neither single-qubit nor two-qubit operations change the initial state $\ket{\downarrow}^{\otimes N}$, since $X(0)$ is a microwave drive of duration $t_{\rm MW} = 0$, and the CNOT logic remains inactive when the qubits remain in $\ket{\downarrow}$. However, since the CNOT gates are compiled using a sequence of single-qubit rotations and free precession under the exchange interaction (Fig.\,\ref{fig:Figure2}\,c), we can expect that both gate errors and dephasing contribute to a reduction in visibility even at $\theta = 0$. Because the number of CNOT gates increases with system size, compounding errors could explain why the observed return probability at $\theta=0$ drops with $N$.

When $\theta\neq n\pi$ with $n\in\mathbb{N}$, the operation $X(\theta)$ brings the qubits involved to a point in the Bloch sphere where they become susceptible to dephasing. An important observation is that when gates are implemented sequentially, as is the case here, dephasing effects quickly accumulate as the circuit depth increases due to the many idling qubits. In some cases, the idling times experienced by specific qubits become comparable to their $T_2^*$ times, as seen in Fig.\,\ref{fig:Figure3} and discussed further below. Additionally, the $X(\theta)$ rotations themselves introduce small errors, which result in a lower return probability for $\theta = 2\pi$ than for $\theta=0$. This is aggravated by qubit dephasing following imperfect $X(2\pi)$ rotations and imperfect CNOTs.

Aimed at modeling the impact of dephasing, we quantify $T_2^*$ using Ramsey-like experiments in Fig.\,\ref{fig:Figure3}\,a. We note that spin relaxation is typically negligible compared to dephasing due to its timescale being orders of magnitude longer than those of interest ($\approx$ 100's ms). 
The dephasing rates have been shown to be contingent on whether the qubits are operated in isolation or with active exchange interaction $J$\,\cite{xue2022quantum}. For this reason, we present $T_2^*$ measurements on all qubits in the two-qubit subspace over a range of barrier voltages taking the coupling from `on' to `off'. Dephasing is mostly caused by charge noise displacing the electrons in the micromagnet magnetic field gradient\,\cite{paquelet2023reducing}, as well as by residual $^{29}$Si nuclear spins coupling through the hyperfine interaction. In the $J_{\rm{on}}$ regime, the qubit frequency additionally becomes sensitive to charge noise modulating the exchange interaction. Despite operating at the symmetry point (see Fig.\,S6 from Supplementary Material S7), where the impact of charge noise is expected to be minimized\,\cite{reed2016reduced}, we indeed observe a reduction in $T_2^*$ times as $J$ increases for all qubit pairs. Interestingly, the dephasing times for $J_{\rm{off}}$ remain close to the values measured over two years earlier on the same device \,\cite{philips2022universal}, even though we now operate at 200 mK and multiple thermal cycles to room temperature have taken place.

We compare the measured $T_2^*$ values to the idling times for each qubit as a function of varying system size $N$ and circuit length $\theta = \{0,\pi/2,\pi,2\pi\}$ (excluding initialization and readout) in Fig.\,\ref{fig:Figure3}\,b. The idling periods are indicated by the yellow  lines connecting the quantum gates in Fig.\,\ref{fig:Figure2}\,a. The horizontal dashed lines in Fig.\,\ref{fig:Figure3}\,b depict the $T_2^*$ times obtained from Fig.\,\ref{fig:Figure3}\,a in the $J_{\rm{off}}$ and $J_{\rm{on}}$ regimes. Here, $T_2^*$ in the $J_{\rm{on}}$ regime is the average $T_2^*$ calculated across all possible interaction conditions. For example, for qubit 3, we take the mean value for the $T_2^*$ obtained when qubit 2 is in either the $\ket{\downarrow}$ or $\ket{\uparrow}$ state, as well as when qubit 4 is in either the $\ket{\downarrow}$ or  $\ket{\uparrow}$ state. Notably, for some qubits such as qubit 1, 2 and 5, the idling times in the circuit exceed their respective $T_2^*$. 

Using the $T_2^*$ measurements, we model the evolution of the $N$-qubit state in the presence of quasi-static phase noise during both idling times and quantum operations. The details can be found in Supplementary Section S8, and the results are presented in Fig.\,\ref{fig:Figure2}\,f.  As in the experimental plots, we show the average behavior obtained when running the simulations for all possible qubit combinations, using actual gate durations and measured $T_{2}^{*}$ times. While this model is not intended to quantitatively reproduce the measured data, it is sufficient to observe that pure dephasing competing with the operational speed can reproduce some of the trends present in the data, such as an $N$-dependent reduction in the visibility of the oscillations (including at $\theta=0$) and a decay as a function of $\theta$. 

The Loschmidt echo amplitude only captures how dissimilar the evolved quantum state is from the initial state, offering no additional information about the time-evolved state itself. However, for specific qubit combinations (i.e. 2–3–4; 4–3–2; 3–4–5; 5–4–3; 2–3–4–5 and 5–4–3–2), it is also possible to access the value of the magnetization along the $z$ axis, $M_z = \frac{1}{N}\sum_{i=1}^NZ_{i}$, using the readout sequence of Fig.~\ref{fig:Figure1}\,a. Specifically, we can directly exctract a measure of $Z$ for qubits 3 and 4 from $\mathcal{M}_{\rm{QND}}$, and for qubits 2 and 5 through $\mathcal{M}_{ZZ}^{\rm{C}}$, provided that neither qubits 1 and 6 are involved in the circuit. The magnetization along the $z$-axis is expected to behave as shown in  Fig.\,\ref{fig:Figure2}\,g, reaching zero at $\theta_{\rm{c}}$, which marks the point at which the state is anticipated to be maximally entangled. Fig.\,\ref{fig:Figure2}\,h displays the experimentally obtained average magnetization for the accessible qubit combinations as a function of $\theta$. In order to measure the reduced set of observables needed to extract the magnetization, we leave out the steps (CNOT+$\mathcal{M}_{ZZ}^{\rm{D}}$) from the readout sequence. Notably, this contributes to a lower loss of visibility for Fig.\,\ref{fig:Figure2}\,h than seen in Fig.\,\ref{fig:Figure2}\,e. We further compare the experimental results with numerical simulations which incorporate quasi-static dephasing noise shown in Fig.\,\ref{fig:Figure2}\,i, which capture some of the deviations from the ideal result seen in experiments.

\subsection{\label{sec:state_tomo}Quantum state tomography}

\begin{figure*}[!t]
\includegraphics[width=0.9\textwidth]{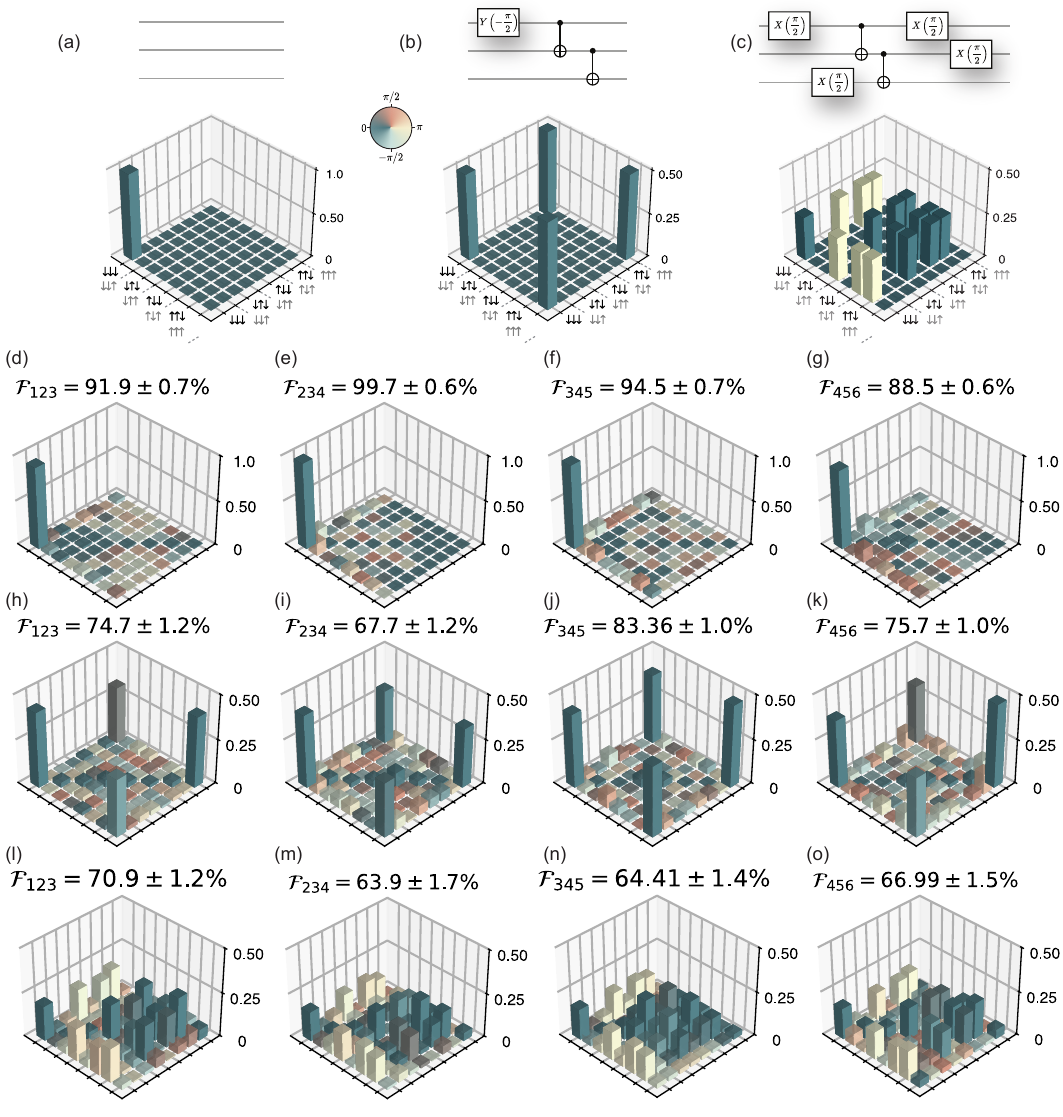}
  \caption{\textbf{Quantum state tomography for $N=3$ at 200 mK.} \textbf{(a-c)} Quantum circuits and expected density matrices for \textbf{(a)} the $\ket{\downarrow\downarrow\downarrow}$ state, \textbf{(b)} a GHZ state and \textbf{(c)} the state obtained when executing the quantum circuit for $N=3$ and $\theta$ = $\pi$/2. The initialization and tomographic pulse sequence for state readout are omitted in the quantum circuits. \textbf{(d-g)} Experimentally measured density matrices for the circuit in \textbf{(a)}, for four sets of three qubits $ijk$ as indicated. The state fidelities $\mathcal{F}_{ijk}$ shown above each panel are calculated compensating for readout errors, and the error bars represent one standard deviation ($1\sigma$), calculated from 300 bootstrap resampling iterations. \textbf{(h-k)} Experimentally measured density matrices for the circuit in \textbf{(b)}. The state fidelities for the GHZ states are calculated without loss of generality by considering the general form $\ket{\psi_{\rm{GHZ}}}$ = ($\ket{\downarrow\downarrow\downarrow}$ + $e^{i\phi} \ket{\uparrow\uparrow\uparrow}$)/$\sqrt{2}$ and fitting the arbitrary phase $\phi$ to maximize the fidelity. The analysis yields the modest values $\phi_{123}$ = -0.54, $\phi_{234}$ = 0.16, $\phi_{345}$ = 0.03, $\phi_{456}$ = -0.54. 
  \textbf{(l-o)} Experimentally measured density matrices for the circuit in \textbf{(c)}. The raw state fidelities without removal of readout errors are \textbf{(b)} 84.73 $\%$ \textbf{(c)} 87.08 $\%$ \textbf{(d)} 87.01 $\%$ \textbf{(e)} 87.87 $\%$ \textbf{(g)} 59.40 $\%$ \textbf{(h)} 47.89 $\%$ \textbf{(i)} 55.38 $\%$ \textbf{(j)} 57.71 $\%$ \textbf{(l)} 61.26 $\%$ \textbf{(m)} 47.90 $\%$ \textbf{(n)} 46.23 $\%$ \textbf{(o)} 59.43 $\%$.}
  \label{fig:Figure4}
\end{figure*}

The Loschmidt echo and magnetization are natural observables in the context of dynamical quantum phase transitions. As a means for assessing in detail how well the full system behaves under realistic workloads, we next perform full quantum state tomography (QST). This gives insight, for instance, in the overall fidelity of the state at relevant points in the quantum circuit, and in how well entanglement spreads across the array. We choose to examine three states, presented in Fig\,\ref{fig:Figure4}; (a) the input state $\ket{\downarrow\downarrow\downarrow}$, (b) a three-qubit Greenberger-Horne-Zeilinger (GHZ) state for comparison, and (c) the final state when running the circuit in Fig.\,\ref{fig:Figure2}\,\,a for $N = 3$ and for the critical angle $\theta_{\rm c} = \pi/2$, the angle $\theta$ that is expected to produce the strongest entanglement. The state $\ket{\psi(\theta_{\rm c})}$ is given by $\frac{1}{4}\left(\ket{\downarrow\downarrow\downarrow}+\ket{\downarrow\uparrow\uparrow}+\ket{\uparrow\downarrow\uparrow}+\ket{\uparrow\uparrow\downarrow}\right)$.
    
The set of operators used for the tomographic pulses  are identical to those reported in \,\cite{philips2022universal}, and we employ Maximum Likelihood Estimation (MLE) to obtain the density matrix $\rho_{\rm{exp}} $. We limit the tomography to three and four qubits, as expanding the measurement space further is constrained by the coherence of the system. For example, the maximum duration of the pulse sequences required to construct the observables for tomography on four qubits already ranges from 830 to 1227 ns, depending on the specific qubits involved, with an average sequence duration of 560 to 745 ns. For five qubits, the range is 1405 to 1347 ns, with an average sequence length of 830 

\clearpage
to 860 ns, while experiments for six qubits require pulse sequences up to 1922 ns long, with a mean of 1134 ns. These estimates only consider the duration of the gates used to construct the tomographic operators, excluding any additional time required for preparing the state of interest.
Fig. \ref{fig:Figure4} displays the quantum state tomography results for various subsets of three qubits along the array (data for four qubits are presented in Supplementary S9). For reference, subfigures (a, b, c) depict the ideal target state for the respective quantum circuits. At the top of each panel showing experimental data, we indicate the state fidelities $\mathcal F =  \bra{\psi} \rho_{\textrm{exp}} \ket{\psi}$ obtained after measurement error removal, with $\ket{\psi}$ the ideally expected state. The results are consistent with state-of-the-art values reported for semiconductor qubit platforms\,\cite{takeda2021quantum, philips2022universal}. We find that compared to previous work on this device\,\cite{philips2022universal}, the unexpected arbitrary phases present in the off-diagonal elements of the density matrices for the GHZ states are largely absent. A similar improvement is seen in  QST results on Bell States for all pairs of neighbouring qubits (Supplementary Material S9).
We explain this advance compared to earlier results in part by the fact that all the measurements were taken at 200 mK. This minimizes spurious phase accumulation due to microwave crosstalk, and allows us to avoid pre-pulsing to saturate and thus stabilize the spin resonance frequencies before operation\,\cite{undseth2023hotter}.

\section{\label{sec:Discussion}Discussion}
In conclusion, we demonstrated the implementation of a realistic quantum circuit on a semiconductor spin-qubit device, utilizing a qubit register of up to six qubits—representing the largest quantum circuit executed in semiconductors to date. The circuit was inspired by a protocol designed to investigate dynamical quantum phase transitions in Ising-like systems. Its scaling features allowed us to benchmark the performance of our quantum processor as a function of system size $N$. By measuring the return probability and magnetization, we recover the expected periodic dynamics of the circuit, demonstrating that controlled interactions, enabled by sequentially activating neighboring couplings, can be achieved across the entire array.
We find that the performance of the quantum processor is limited by dephasing, in particular for idling qubits, as well as by errors in state preparation and measurement. To assess the effects of dephasing, we experimentally measure $T_2^*$ for each qubit, and numerically simulate the time evolution of the system accounting for quasi-static phase noise both when idling and during single-qubit rotations. The numerical simulations qualitatively reproduce the main features observed in the experimental data, such as a reduction in visibility as a function of $N$ as well as a degradation in the signal amplitude for longer evolution times. An additional overall reduction in the experimental visibility results from SPAM errors. 

Unlike previous approaches that separately benchmark single- and two-qubit gates for individual qubits or qubit pairs, or independently assess state preparation and measurement errors, the present implementation takes a more holistic approach. In all cases, we initialize and perform readout on all six qubits, and execute qubit operations as a function of a variable parameter $\theta$. This method enables us to evaluate the processor’s capabilities comprehensively, without confining the analysis to specific quantum states or benchmarks and compare the results with simulations. We further characterize  the overall system performance through quantum state tomography, and highlight the importance of improving SPAM errors for better state estimation.

Future work should focus on mitigating dephasing effects from idling through the parallelization of single-\,\cite{lawrie2023simultaneous} and two-qubit operations and possibly also by adopting new control strategies that achieve faster qubit operations, such as optimized baseband single-qubit control\,\cite{unseld2024baseband}. In addition, it will be important to enhance dephasing times, for instance through optimized nanomagnets that suppress decoherence gradients\,\cite{aldeghi2024simulation}, and by reducing charge noise with improved dielectric gate stacks and using thinner quantum wells\,\cite{paquelet2023reducing}. To further reduce hyperfine noise, isotopic purification should be improved compared to the current residual 800ppm spinful nuclei and extended to the SiGe barrier above the quantum well.

\section{\label{sec:Acknowledgements}Acknowledgements}
We acknowledge the contribution of A. Sammak to the development of the $^{28}$Si/SiGe heterostructure. We thank M. Rimbach-Russ for input regarding the theoretical models, K. Capanelli for assistance with the data analysis rethresholding method and members of the Vandersypen groups for useful discussions. We acknowledge
financial support from Intel, and the Army Research Office (ARO) under grant numbers W911NF-17-1-0274 and W911NF-12-1-0607. The views and conclusions contained in this document are those of the authors and should not be interpreted as representing the official policies, either expressed or implied, of the ARO or the US Government. The US Government is authorized to reproduce and distribute reprints for government purposes notwithstanding any copyright notation herein. Development and maintenance of the growth facilities used for fabricating samples is supported by DOE (DE-FG02-03ER46028). We further acknowledge the financial support from the Dutch Ministry for Economic Affairs through the allowance for Topconsortia
for Knowledge and Innovation (TKI) as well as support from the “Quantum Inspire–the Dutch Quantum Computer in the Cloud” project (Project No. NWA.1292.19.194) of the NWA research program “Research on Routes by Consortia (ORC),”  funded
by the Dutch Research Council (NWO).
\section{\label{sec:data_availability}Data Availability Statement} 

Data and scripts used in this publication are available in the Zenodo repository https://doi.org/10.5281/zenodo.15480420.
\section{\label{sec:Methods}Methods}
\subsection{Parity-mode Pauli Spin Blockade at the (3-1)-(4-0) charge configuration.}
Spin-to-charge conversion occurs at the (3,1)-(4,0) charge transition. In the (3,1) configuration, two electrons occupy the lowest energy valley state and one electron fills the next available valley orbital in the first dot, while the second dot contains one electron occupying the lowest valley state. Compared to the (1,1) occupation, this results in a larger energy gap to the next available orbital for the moving electron (from dot 2 or 5 in this case), thus increasing the readout window\,\cite{philips2022universal}. The Zeeman energy difference between qubits enables operation in the parity-mode regime, where the $\ket{T_0}$ state rapidly relaxes to the singlet state during the readout window, while the $\ket{T_+}$ and $\ket{T_-}$ states remain blockaded\,\cite{seedhouse2021pauli}.  

\subsection{Zeeman energy differences and CZ gates.}

In the current settings, the Zeeman energy difference, $\Delta E_{\rm Z}$, between neighboring qubits is around 100's MHz, and exchange is operated in the few MHz regime. The comparatively large Zeeman energy difference suppresses the flip-flop terms in the exchange interaction, leading to a $ZZ$ coupling Hamiltonian, which naturally realizes the CZ gate. An exception to this are qubits 3 and 4 which have a $\Delta E_{\rm{Z}}$ of 12 MHz. However, adiabaticity in the switching of the $J$ interaction ensures that the two-qubit operation is still a CZ gate~\cite{rimbach2023simple}.

\subsection{Calibration routines} 
\label{sec:calibrations}
The measurement routines employed for the calibration of the device, were the same as those discussed in detail in  Figure 4 of the  Extended Data from the work of Philips et al.\cite{philips2022universal}.
Here, operating at 200mK allowed us to bypass the single-qubit phase crosstalk calibrations shown in subfigure \textbf{c}, as crosstalk from microwave heating was mitigated. 

\subsection{CNOT decomposition using the CZ gate}
We leverage the native $ZZ$ interaction to construct the CNOT operation, using the decomposition  CNOT$:= \mathrm{H}\, \mathrm{CZ}\, \mathrm{H}$, where H is the Hadamard gate and  $\mathrm{CZ}:= Z_1(-\pi/2)Z_2(-\pi/2)U_{\rm{ZZ}}(\pi\hbar/J)$, where $U_{\mathrm{C}Z}$ is the unitary operator describing the $ZZ$ interaction, $\hbar$ is the reduced Plank's constant and $J$ is the value of the exchange interaction strength. The argument of the operator $U_{\mathrm{C}Z}$ corresponds to the interaction time needed to obtain $\mathrm{C}Z= \rm{diag}(1,1,1,-1)$.
In our implementation of the CNOT, the Hadamard gates are substituted by $Y(-\pi/2)$ and $Y(\pi/2)$. The equivalence is rooted in the fact that the Hadamard gate can be decomposed as $H=Y(-\pi/2)Z$ and $H=ZY(\pi/2)$. Since the $Z$ gates commute with the CZ operation, they cancel out, resulting in the $Y$-based formulation.
\onecolumngrid
\newpage

\pagebreak


\twocolumngrid
%

\end{document}


\onecolumngrid
\setcounter{equation}{0}  
\begin{center}
\textbf{\huge Supplementary Material}
\end{center}
\bigskip
This Supplementary Material includes:
\begin{itemize}
    \item S1. Rethresholding of readout signal\dotfill\pageref{sec:rethresholding} 
    \item S2. Rabi experiments \dotfill\pageref{sec:rabis}
    \item S3. Biased probability after logical-AND operation \dotfill\pageref{sec:bias_and_operation}
    \item S4. Analytical formulation of the circuit dynamics\dotfill \pageref{sec:simplification xx interaction}
    \item S5. Effect of two-qubit interaction in the periodicity \dotfill \pageref{sec:periodicity-twoqubit}
    \item S6. Traces for all possible qubit combinations in the quantum circuit \dotfill \pageref{sec:all_traces}
    \item S7. Characterization of the exchange interaction \dotfill \pageref{sec:exchange}
    \item S8. Numerical simulations with low-frequency noise \dotfill \pageref{sec:numericalsimulations}
    \item S9. Quantum state tomography \dotfill \pageref{sec:Bell_states}
\end{itemize}
\clearpage
\section{Rethresholding of readout signal}
\label{sec:rethresholding}
\begin{figure*}[ht]
  \centering
\includegraphics
[width=\textwidth]{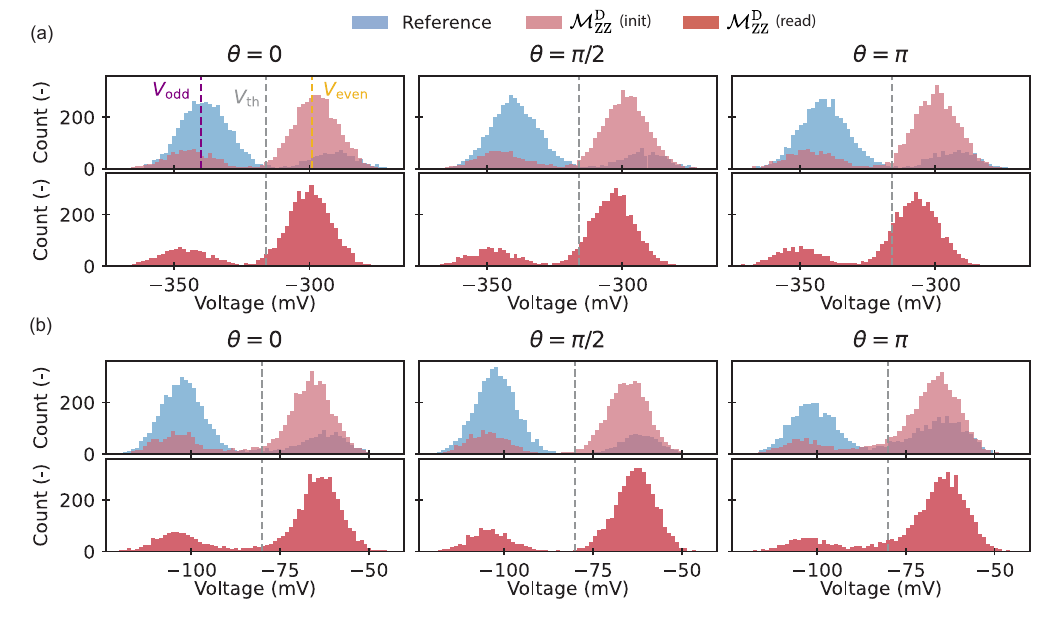}
  \caption{\textbf{Drift of sensing dots from microwave heating.} 
  Histograms obtained by binning 5000 PSB readout signals (shots) collected while performing $X(\theta)$ rotations on \textbf{(a)} qubit 1 and \textbf{(b)} qubit 6. The microwave burst durations correspond to rotation angles of $\theta$ = 0, $\theta$ = $\pi$/2, and $\theta$ = $\pi$. Representative traces of the readout signal for each $\theta$ are shown for distinct steps during the initialization (first row \textbf{(a-b)}) and readout (second row \textbf{(a-b)}) sequences presented in  Fig.\,1\,a. The histogram labeled `Reference' corresponds to the first PSB measurement $\mathcal{M}_{ZZ}^{\rm{A}}$ in the initialization sequence, before any microwave pulsing occurs. The histograms labeled $\mathcal{M}_{ZZ}^{\rm{D}}$(init) and $\mathcal{M}_{ZZ}^{\rm{D}}$(read) correspond to the measurement $\mathcal{M}_{ZZ}^{\rm{D}}$ in the initialization and readout sequence, respectively.  
 The grey dashed lines represent the fixed voltage threshold $V_{\rm{th}}$, as determined through independent calibrations of the sensor when no dynamical re-thresholding is applied. An optimal threshold effectively differentiates between the voltage distributions corresponding to the odd ($V_{\rm{odd}}$) and even ($V_{\rm{even}}$) states.
 }

  \label{fig:ext:SDdrift}
\end{figure*}
As explained in Section\,II and the Methods Section, spin-to-charge conversion is achieved through parity-mode Pauli Spin Blockade by correlating the charge state of a pair of electron spins in a double-dot system with their spin parity. In this work, the charge states (3,1) and (4,0) are attributed to even or odd spin parity, respectively. The measurement apparatus is a nearby larger quantum dot operated with RF reflectometry\,\cite{liu2021radio}, which serves as a charge probe for the electron pair. Readout is achieved by integrating the voltage of the reflected RF signal for $\approx 20$ \textmu{}s. Under ideal conditions, the measured voltage assumes one of two distinct values, which directly correspond to the parity state of the spin pair. In practice, noise introduces uncertainty causing these two discrete values to spread into two Gaussian distributions, centered around  $V_{\rm{even}}$ and $V_{\rm{odd}}$, as shown in the histograms of Fig.\,\ref{fig:ext:SDdrift}.

The task at hand is then to find the optimal voltage value $V_{\rm{th}}$ that best distinguishes between these two distributions, enabling us to sort the results accordingly. Ideally, this value remains constant for the entirety of the experiment, here marked with the gray dashed lines in Fig.\,\ref{fig:ext:SDdrift}.

\begin{figure*}[h]
  \centering
\includegraphics[width=\textwidth]{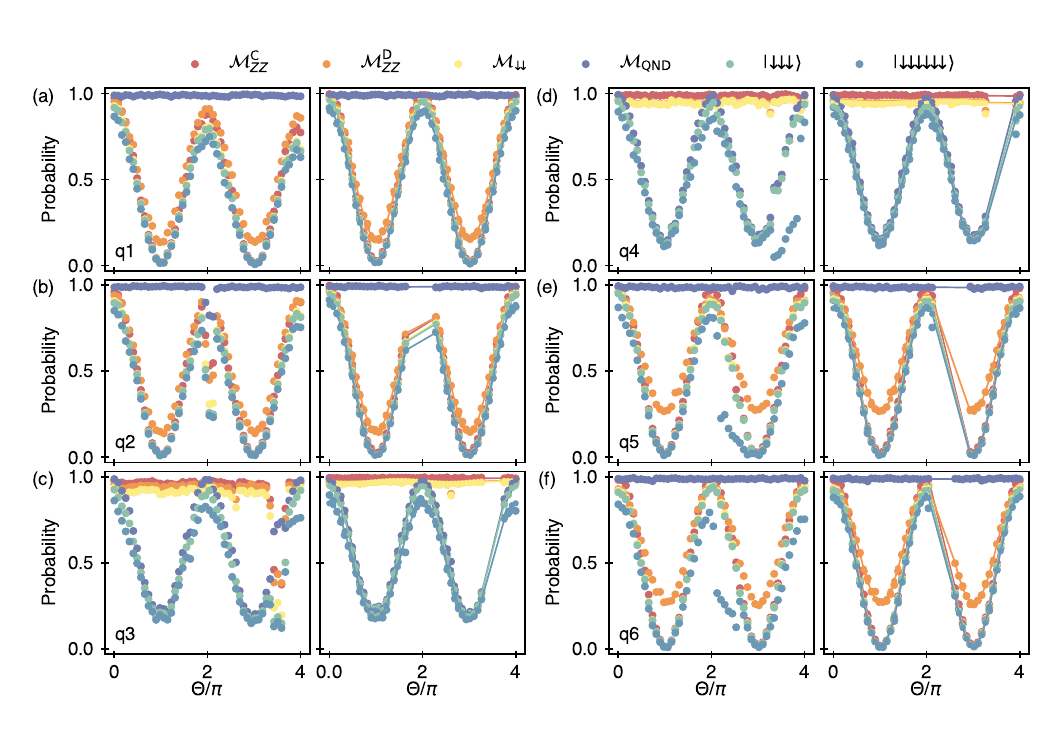}
  \caption{\textbf{Rabi oscillations and effect of retresholding the sensing signal} for qubit 1 \textbf{(a)}, qubit 2 \textbf{(b)}, qubit 3 \textbf{(c)}, qubit 4 \textbf{(d)}, qubit 5 \textbf{(e)} and qubit 6 \textbf{(f)}. For all cases, the state of the entire array is initialized and measured in the $\ket{\downarrow}^{\otimes 6}$ basis, independent of the specific qubit being targeted. However, the plots exclusively display the intermediate readouts from their associated sensor (SD1 for qubits 1, 2 and 3 and SD2 for qubits 4, 5 and 6). The left panels of each subplot display the data measured using a fixed threshold voltage during readout. Due to drifts from microwave heating, a reduction in visibility is recorded as $\theta$ increases. This effect is less pronounced for qubits 4, 5 and 6, implying that SD 2 exhibits a reduced sensitivity to microwave heating. By implementing dynamical re-thresholding during post-processing, as explained in Section\,\ref{sec:rethresholding}, we obtain the panels on the right of every subfigure. More noticeably for qubits 1, 2 and 3, the visibility is more consistent throughout the entire experiment.
A few points in the collected data could not be corrected using the re-thresholding method. Specifically, this pertains to instances where the sensor experiences \textit{``jumps''}, making it impossible to distinguish clearly the two characteristic Gaussian distributions obtained during readout as depicted in Fig.\,\ref{fig:ext:SDdrift}. In the re-thresholded plots, these points were removed. This is also the reason why some datapoints are missing in Fig.\,1h. During other measurements shown in the main text, no sensor jumps occurred.}
  \label{fig:ext:rabis}
\end{figure*}

During device operation, we observe that applying microwave bursts induces a transient shift in the sensor signal, displacing the histograms from their original position and degrading our ability to distinguish the charge states with a constant threshold. Furthermore, we find that the shift increases with the duration of the microwave burst, evidenced by the signal decay of the Rabi experiments in Fig.\,\ref{fig:ext:rabis}. This effect was previously observed~\cite{undseth2023hotter}. 

To illustrate the impact on the present experiment, in Fig.\,\ref{fig:ext:SDdrift} we show multiple readout traces extracted during Rabi experiments. The histograms compile 5000 measurements (referred to as shots) recorded during readout for three different durations of a microwave burst noted as $\theta$ = 0, $\theta$ = $\pi$/2 and $\theta$ = $\pi$ depending on the angle of rotation, where $\theta=\pi$ lasts on the order of hundreds of nanoseconds. We note that a wait time of around 1\textmu{}s is added at the end of each shot, which allows the sensor to settle back to its stable condition before the beginning of the next shot. In the top row of both panels (a) and (b) of the figure, we see that already during the initialization sequence, which contains a microwave burst itself, a shift in the sensor signal histograms is present. As can be expected, this shift does not depend on $\theta$ as the rotation over a variable angle $\theta$ occurs only after initialization. An additional $\theta$-dependent shift is seen in the bottom rows of the figure panels.

To correct for the displacement of the threshold voltage, we individually fit the histogram of each readout trace to a bimodal Gaussian function $f(V) =  A_1\exp(-(V-V_{\rm{odd}})^2/2\sigma_{\rm{odd}}^2)+A_2\exp(-(V-V_{\rm{even}})^2/2\sigma_{\rm{even}}^2)$ and set the threshold at the voltage where the two fitted Guassians show minimal overlap. We find that using this method is an effective way for restoring an optimal readout, with no measurement overhead. We note that the fit relies on the presence of two Gaussian distributions in the readout signal, which will be state and error dependent. 

\newpage
\section{Rabi experiments}
\label{sec:rabis}

\newpage
\section{Biased probability after logical-AND operation}\label{sec:bias_and_operation}
The scheme calls for recording the $\ket{\downarrow}^{\otimes N}$ state. This probability is extracted through AND operations on several readout outcomes. In the presence of readout errors, the AND operation biases the outcome toward lower values, as can be simply understood as follows. Assuming the pre-measurement state is actually $\ket{\downarrow}^{\otimes N}$, we define the probability of obtaining an outcome of 1 on the first PSB measurement as
\begin{equation}
    P\left(\rm{PSB1} = 1\right) =  p.
\end{equation}
Likewise for the outcome 1 of the second PSB measurement we have 
\begin{equation}
    P\left(\rm{PSB2} = 1\right) =  q
\end{equation}
The probability of measuring 1 for both readouts is given
\begin{equation}
    P_{11} = P\left(\rm{PSB1} = 1 \ \& \ \rm{PSB2} = 1\right) =  p\cdot q.
\end{equation}
If $p<1$ and $q<1$, then it follows that
\begin{equation}
    p\cdot q < \min(p, q) \;.
\end{equation}

\newpage
\section{\label{sec:simplification xx interaction} Analytical Formulation of the Circuit Dynamics}
\subsection{Boundary conditions}
The existence of the finite DQPT-like behavior for mutually-commuting Pauli operator models or so-called stabilizer models is entirely dependent on the existence of Pauli operator \emph{constraints} as argued in \cite{Schmitz2020}. A constraint in this context is any product of Hamiltonian terms whose product is proportional to the identity.  Whereas such constraints occur naturally for the 1D Ising model on a cycle, the quantum circuit implementation presented in the main body of the text must add boundary terms to generate a constraint via one-body terms on both sides of the linear array.
Without the constraint, each Pauli operator term in the Hamiltonian acts as an independent spin which is then reflected in the dynamics as such. However, the existence of the constraint requires correlation between terms in the Hamiltonian, even as all terms can be simultaneously diagonalized. Ref.~\cite{Schmitz2020} argues the existence of a dynamical version of the Wegner duality\cite{Wegner1971}, whereby the dynamics of a stabilizer Hamiltonian is equivalent to a dynamical model of its constraints. For the case of our 1D Ising model, the single constraint is given by the product of all the bulk two-body terms and the pair of one-body boundary terms. As this constraint is extensive in the size of the system, we have divergent dynamics in the infinite size limit, in particular at the critical time $\theta_c$ .

\subsection{Circuit Simplification}
In Section III, we explained that the circuit proposed for this work resembles the dynamics of a quantum quench, whereby an initial state given by the ground state of a Hamiltonian $H_0 =h\sum_{i =1}^{N}Z_i$ is taken out of equilibrium by the action of the interaction Hamiltonian $H_{\rm quench}(h = 0) = J\sum_{i=1}^{N-1}X_{i}X_{i+1}$. The unitary describing the time evolution following the quench, can be expressed in terms of single- and two-qubit primitives as
\begin{equation}
    \mathcal{U}_{XX}\left(\theta\right)= \prod_{i = N}^{2} U_{\rm CNOT}(i-1,i)\cdot \prod_{i = 1}^{N-1}R_{x}^{i}\left(\theta\right)\cdot \prod_{i = 1}^{N-1} U_{\rm CNOT}(i,i+1)\cdot R_{x}^{N}\left(\theta\right)\cdot R_{x}^{1} \left(\theta\right),
    \label{Eq:DQPT_unitary_full}
\end{equation}
where $R_{x}^{i}\left(\theta\right) 
$ represents a single qubit rotation around the $x-$axis by an angle $\theta$,  and where $U_{\rm CNOT}(i,i+1)$ denotes a Controlled-NOT gate acting on the qubit pair $\{i,i+1\}$, both defined in an $2^{N}\otimes 2^{N}$ dimensional space. Here $\theta$ serves as a time parameter, representing the dynamical component of the circuit.

Following circuit optimization methods found in \cite{Verteletskyi2020, Gokhale2020,Crawford2021,Schmitz2023}, the Clifford sub-circuit denoted as Extended circuit in Fig.\,2\,b can be replaced with a different Clifford circuit. In our case, we propose to replace the CNOT staircase with a trivial circuit. This alternative circuit is then given by

\begin{equation}
    \tilde{\mathcal{U}}_{XX}\left(\theta\right)= \prod_{i = 1}^{N-1}R_{x}^{i}\left(\theta\right)\cdot \prod_{i = 1}^{N-1} U_{\rm CNOT}(i,i+1)\cdot R_{x}^{N}\left(\theta\right)\cdot R_{x}^{1} \left(\theta\right).
    \label{Eq:DQPT_unitary_simplified}
\end{equation}
This equivalence holds only for the native measurement operators in this system and is valid specifically for the initial state $\psi = \ket{\downarrow}^{\otimes N}$.

\subsection{Finite-Size Effects in Loschmidt Echo Amplitude Scaling}
In the context of Dynamical Quantum Phase Transitions\,\cite{heyl2018dynamical}, it is expected that in the thermodynamical limit ($N\rightarrow\infty$), the dynamical free energy density $\lambda = -\rm{ln}(\mathcal{L})/N\rightarrow\infty$ becomes non-analytic at critical times. That condition is satisfied if $\mathcal{L}\left(\theta_{\rm c}\right) = 0$. 
Using the unitary operators of Eq.~\ref{Eq:DQPT_unitary_simplified}, we derive a general analytical expression for the Loschmidt echo as a function of $N$ and $\theta$,
\begin{align}
   \mathcal{L}\left(\theta\right) &= \left|\bra{\downarrow}^{\otimes N}\mathcal{U}_{XX}\left(\theta\right)\ket{\downarrow}^{\otimes N} \right|^{2} \\ &= \left|\bra{\downarrow}^{\otimes N}\tilde{\mathcal{U}}_{XX}\left(\theta\right)\ket{\downarrow}^{\otimes N} \right|^{2} \\ &= \left|\cos\left(\frac{\theta}{2}\right)^{N+1}+i^{N+1} \sin\left(\frac{\theta}{2}\right)^{N+1}\right|^{2}.
    \label{Eq:loschmidt_analytical}
\end{align}

\begin{figure*}[h]
  \centering
\includegraphics{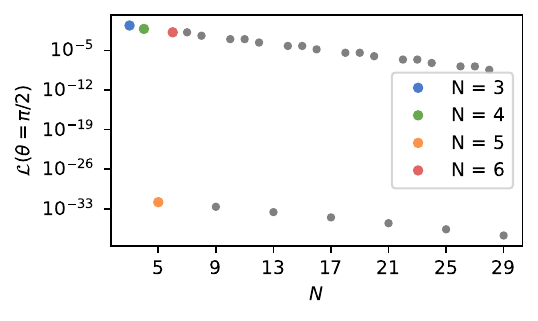}
  \caption{\textbf{Loschmidt echo amplitude at critical angle $\theta_{\rm c}=\dfrac{\pi}{2}$ vs system size $N$.}}
  \label{fig:Supp_finite_size}
\end{figure*}

From Eq.\,\ref{Eq:loschmidt_analytical} it follows that the Loschmidt echo approaches 0 at a rate that depends on $N$, generally increasing as $N$ becomes larger. However, finite size effects introduce exceptions to this trend. Figure\,\ref{fig:Supp_finite_size} shows the Loschmidt echo amplitude on a logarithmic scale evaluated at $\theta_{\rm c} = \pi/2$, where minima are expected. Rather than a continuous trend towards zero, we observe that the minima are approached in an oscillatory manner with a period of $N=4$. This explains why the curve for $N = 5$ on Fig.\,2\,d reaches a lower minimum than for $N = 6$.

\clearpage
\newpage
\section{Effect of two-qubit interactions on the periodicity of the Loschmidt echo}
\label{sec:periodicity-twoqubit}
\begin{figure*}[h!]
  \centering
 \includegraphics[width = 0.9\textwidth]{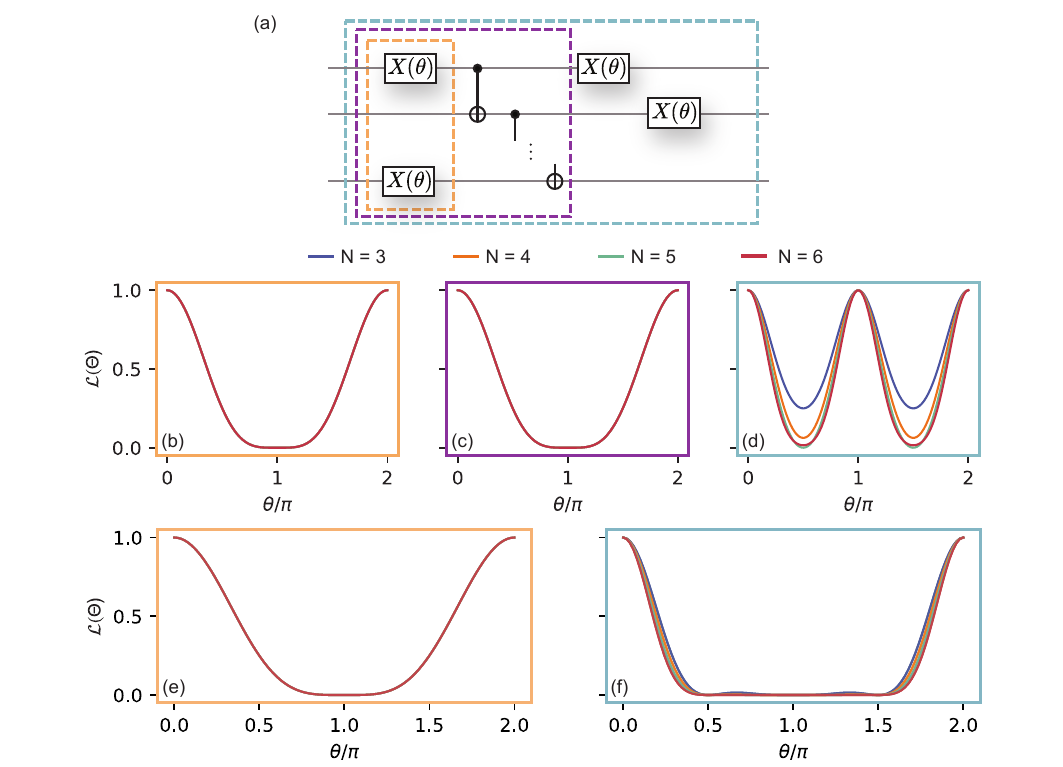}
  \caption{\textbf{Effect of two-qubit interactions on the periodicity of the Loschmidt echo.}. \textbf{(a)} Circuit schematic, with dotted channels indicating the variable size $N$. The dashed boxes highlight key stages during the circuit's execution, where the periodicity of the Loschmidt echo is particularly interesting to analyze. Two distinct periodicity patterns are observed as a function of $\theta$ and $N$. Before \textbf{(b)} and immediately following \textbf{(c)} the entangling gates, the return probability exhibits a periodicity of $\theta = 2\pi$. However, when entangling gates are present, the final stage, highlighted by the turquoise box, reveals oscillations with a periodicity of $\theta = \pi$, and an amplitude that depends on $N$. In contrast, for cases where no entangling gates are present \textbf{(e-f)}, both outcomes retain a periodicity of $\theta = 2\pi$, indicating that the responsible element for the periodicity in the circuit are the two-qubit interactions.}
  \label{fig:Supp_periodicity_sims}
\end{figure*}

\newpage
\section{\label{sec:all_traces}Traces for all possible qubit combinations in the quantum circuit}
\begin{figure*}[ht]
  \centering
\includegraphics[width=.9\textwidth]{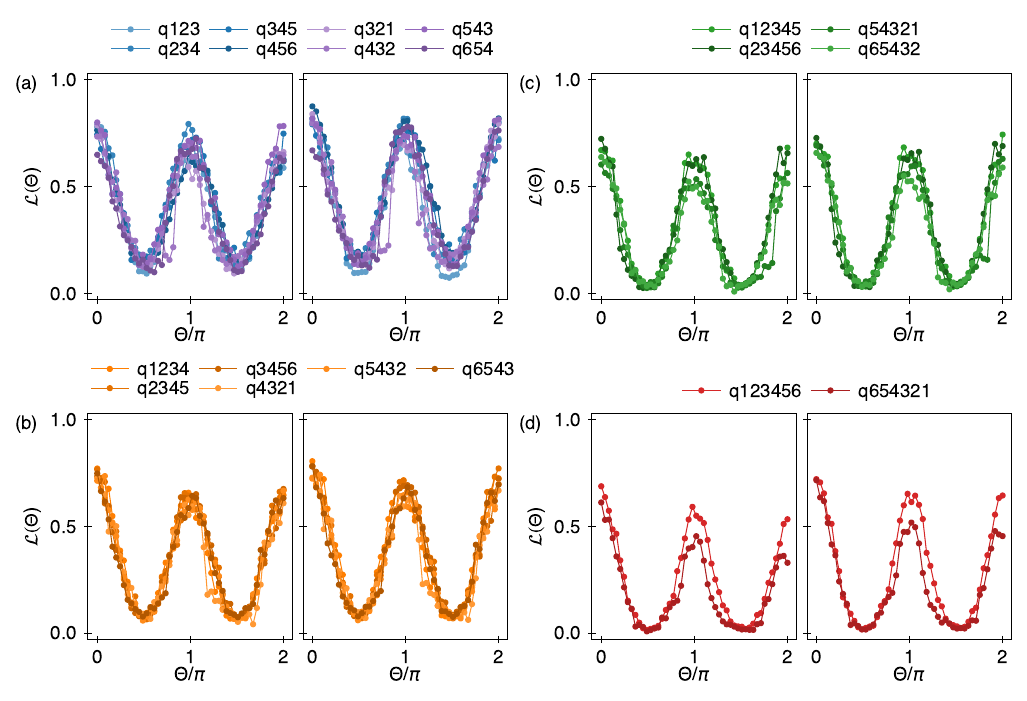}
  \caption{\textbf{Separate experimental data for all qubit combinations.}
  \textbf{(a)-(d)} The Loschmidt echo amplitude is presented for all possible combinations of varying numbers of qubits, as obtained while implementing the circuit described in Fig.\,2\,a of the main text. The subfigures correspond to the experimental data acquired from configurations involving \textbf{(a)} 3 qubits, \textbf{(b)} 4 qubits, \textbf{(c)} 5 qubits, and \textbf{(d)} 6 qubits, respectively. In each subfigure,  the left panel displays the raw data using a fixed threshold for readout and the right panel presents the data after applying dynamic re-thresholding as explained in Section\,\ref{sec:rethresholding}. The data shown in Fig.\,2\,e represents for each value of $N$ the average calculated across all datasets presented in this figure.}
  \label{figsupp:DQPTs_all}
\end{figure*}

\newpage
\section{Characterization of the exchange interaction}
\label{sec:exchange}
\begin{figure*}[h!]
  \centering\includegraphics[width=\textwidth]{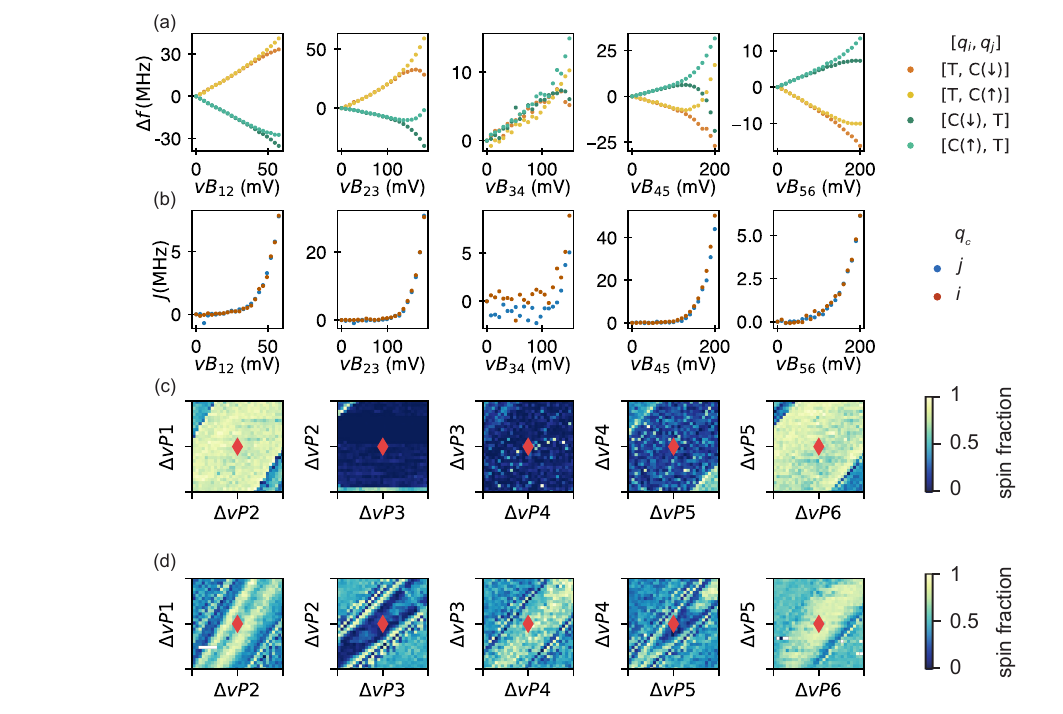}
  \caption{\textbf{Variation of frequency, exchange coupling and symmetry scans.} \textbf{(a)} Frequency detuning as a function of the virtual barrier voltage, $vB_{ij}$, for each qubit pair $(q_i,q_j)$. At low barrier amplitudes, the change in frequency is explained by the displacement of the electrons within the micromagnet field gradient, and a linear trend is observed. At sufficiently high voltages, the exchange interaction causes each spectral line to split into two distinct branches. Each branch corresponds to the resonance frequency of the target qubit T, for a unique state of the control qubit C, prepared in either the $\ket{\uparrow}$ or $\ket{\downarrow}$ state during the experiment. \textbf{(b)} The value of the exchange interaction strength $J$ can be isolated by substracting each branch in \textbf{(a)} and the expected exponential trend as a function of $vB_{ij}$ is obtained for each pair, independently of the qubit, $q_c$, used as the control qubit. \textbf{(c)-(d)} Plots used to verify the position of the symmetry points~\cite{reed2016reduced,martins2016noise} obtained by monitoring the return probability to an initial state, after a decoupled CZ gate, as a function of virtual plunger gate detuning $\Delta vP_i$ and $\Delta vP_j$. The subfigures in \textbf{(c)} were obtained in the regime of low barrier voltage, i.e., in the absence of exchange interaction, while those in \textbf{(d)} were measured for amplitudes where the exchange interaction was active. The two regimes are the same as the ones indicated by $J_{\rm off}$ and $J_{\rm on}$ in Fig.\,3. The $x$- and $y$-axis of each subpanel span $\Delta vP = 80$\,mV and the chosen operation point is marked by the red diamond marker. Note that in these plots, an entire line of symmetry points exists. The actual operating point is chosen such that a suitable value of $J_{\rm on}$ is achieved, which is additionally not too sensitive to the exact position on the ``symmetry line''.}
\label{fig:exchange}
\end{figure*}

\newpage
\section{\label{sec:numericalsimulations}Numerical simulations with low-frequency noise}
Here, we describe and justify the approach used to incorporate incoherent noise into the simulations of the circuit shown in Fig.\,2\,a, whose corresponding results (Fig.\,2\,f,i) were presented alongside the experimental data (Fig.\,2\,e,h).

We consider dephasing as the primary source of incoherent noise limiting the quantum processor performance, with a particular focus on the low spectral end, also referred to as quasi-static noise. This regime accounts for timescales where fluctuations in the Hamiltonian parameters can be considered constant throughout the duration of a single shot ($\approx 100\,\text{'s \textmu s}$). 
The spin system under study is well described by the Hamiltonian
\begin{align}
H &= \frac{g\mu_{\rm B}}{2}\sum_{i}B_{i} \sigma_z+\sum_{i}J_i\vec{\sigma}\cdot\vec{\sigma}\\
&\approx \frac{g\mu_{\rm B}}{2}\sum_{i}B_{i} \sigma_z+\sum_{i}J_i\sigma_z\cdot\sigma_z,
\label{eq:Hamiltonian_spins}
\end{align}
where $g\approx 2$ is the gyromagnetic ratio of the electron spin in Si, $\mu_{\rm B}$ is Bohr’s magneton, and $B_{i}$ is the local magnetic field experienced by each electron, including  the inhomogeneous stray magnetic field created by the micromagnet. The term $J_i$ represents the nearest-neighbor exchange interaction strength, which is electrically tunable via the barrier gates (Fig.\,1\,a) and depends exponentially on the gate voltage (Fig.\,\ref{fig:exchange}\,a-b). As discussed in the Methods, due to the differences in Zeeman energy being much larger than the exchange energy and the adiabatic activation of the exchange coupling, we can treat the interaction as Ising-like and disregard the off-diagonal elements $\sigma_{x}\sigma_{x}$ and $\sigma_{y}\sigma_{y}$.

Notably, in the device discussed here, electrical noise affects both terms in the Hamiltonian of Eq.~\ref{eq:Hamiltonian_spins}. Specifically, fluctuations in the local electric environment can physically displace the electrons within the micromagnet gradient, introducing uncertainty in the Zeeman energies, and  also modulate the exchange interaction strength. Additionally, residual $^{29}$Si nuclear spin fluctuations lead to varying Zeeman energies. All of these contribute to variations in the qubit resonance frequencies, typically characterized by the average $T_2^*$ times. We assume that in every shot of the experiment, the fluctuating single-qubit frequencies are sampled from a Gaussian distribution with width $\sigma_{f} = 1/(\sqrt{2}\pi T_2^*)$. As explained in the main text, we measure $T_2^*$ for each spin with the exchange coupling on and off, and with the other (control) spin in either the spin-up or spin-down state (see Fig.\,3). 

We include in the numerical simulations the quasi-static noise contribution during both idling and driven evolution. Firstly, when one qubit is being rotated over an angle $\theta$, the other qubits are idling for a duration $t_i = \frac{\theta}{\pi} t_\pi$, with $t_\pi$ the duration of a $\pi$ rotation. In the rotating frame of each qubit, the randomly sampled single-qubit frequency deviation $f_\epsilon$ for each spin then translates to a rotation around the $z$-axis over a randomly sampled angle $\theta_\epsilon = 2\pi f_\epsilon t_i$, described by the unitary operator 
\begin{align}
R_z^{i}\left(\theta_{\epsilon}\right) = e^{-i\frac{\theta_{\epsilon}}{2}\sigma_z^i},
\label{Eq:noisy_operator}
\end{align}
where $i$ denotes the qubit index it acts upon. Dephasing of idling qubits while performing a CZ operation on two other qubits is modeled analogously.

Secondly, for the qubit that is being driven, we model the time evolution in the rotating frame by the unitary operator
\begin{align}
R_{\rm{drive}}^{i}\left(\theta_{\epsilon},\theta\right) = e^{-\frac{i}{2}(\theta_{\epsilon}\sigma_z^i+\theta\sigma_x^i)} \;,
\label{Eq:noisy_operator_x}
\end{align}
Given that the Rabi frequencies $f_{\rm{Rabi}}\approx 5\mathrm{MHz}\gg\sigma$, the spin is rotated around an axis close to the $x$-axis, and the effects of dephasing are less pronounced when a qubit is driven than while idling.

\begin{figure*}[h!]
\centering
\includegraphics[width = \textwidth]{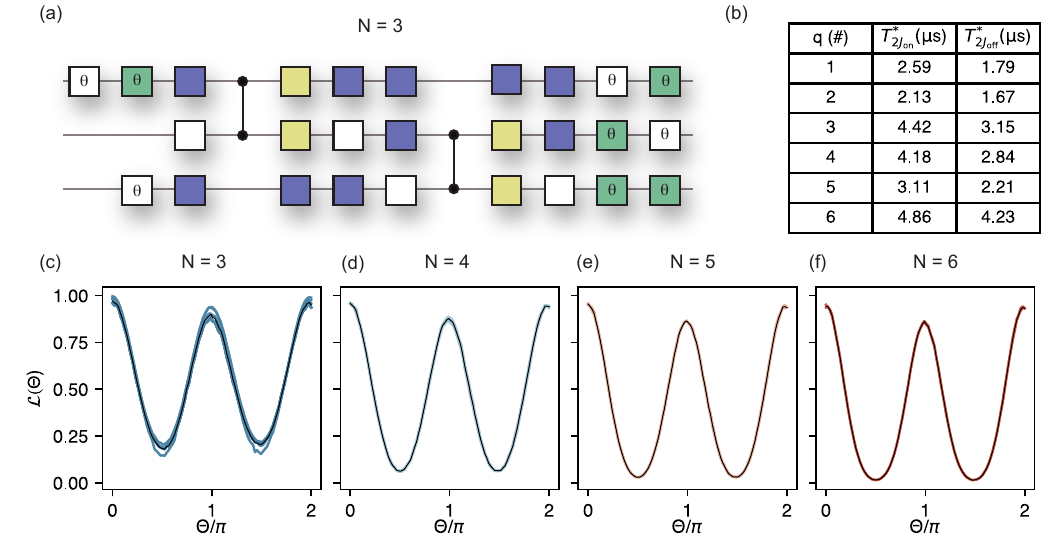}
\caption{ \textbf{Circuit simulations using quasi-stochastic noise}. \textbf{(a)} Example of a three-qubit circuit incorporating dephasing, modeled as single-qubit operators. The white boxes represent noisy single-qubit $X$ rotations, while the colored boxes correspond to noisy $Z$ operations mimicking dephasing during idling (see main text). \textbf{(b)} Table showing the experimentally obtained single-qubit pure dephasing times used in the simulations, measured for both $J_{\rm off}$ and $J_{\rm on}$. \textbf{(c-f)} Numerical simulations of the return probability as a function of $\theta$ for all possible qubit combinations and system sizes.}
\label{fig:simulations_damping}
\end{figure*}

Fig.\,\ref{fig:simulations_damping}\,a schematically summarizes how we model dephasing effects from quasi-static noise for $N=3$. Each colored box represents a step where the time-evolved quantum state is updated by multiplying it by a unitary operator, either $R_z^{i}\left(\theta_{\epsilon}\right)$ or $R_{\rm{drive}}^{i}\left(\theta_{\epsilon},\theta\right)$. In the green and blue boxes, $\theta_{\epsilon}$ is computed based on the duration of the corresponding single-qubit rotation  $\frac{\theta}{\pi} t_{\rm Rabi}$ and on the $T_2^*$ times measured with the exchange interaction off. For the yellow boxes, $\theta_{\epsilon}$ is computed based on the duration of the corresponding CZ gate and on the $T_2^*$ values measured with the exchange interaction switched on. The simulations for $N = 4,5,$ and $6$ follow the same methodology.
The table in Fig.\,\ref{fig:simulations_damping}\,b lists the  $T_2^*$ values used in the simulations for both $J_{\rm off}$ and $J_{\rm on}$. 

The simulation results are presented in Fig.\,\ref{fig:simulations_damping}\,(c)-(f), with each light-colored trace corresponding to a different combination of qubits, while the black curve represents the average of all runs. As with the experimental results, the curves exhibit the expected variability, since the $T_2^*$ times depend on the specific qubits involved in the computation. The noise model based on pure dephasing from quasi-static noise captures some features seen in the experiments (see main text). Importantly, dephasing from quasi-static noise is particularly detrimental when qubits are idling. To achieve closer quantitative agreement between the model and the experiment, the next key step is to account for state preparation and measurement (SPAM) errors. We notice in Fig.\,2 that the discrepancy between the numerical and experimental results is largest near the maxima of the Loschmidt echo. This is exactly what we expect in the presence of measurement errors, as they asymmetrically impact the Loschmidt echo (see Section~\ref{sec:bias_and_operation}).

\label{sec:calibration-errors}

\newpage
\section{Quantum state tomography}
\label{sec:Bell_states}
\begin{figure*}[h!]
  \centering\includegraphics[width=\textwidth]{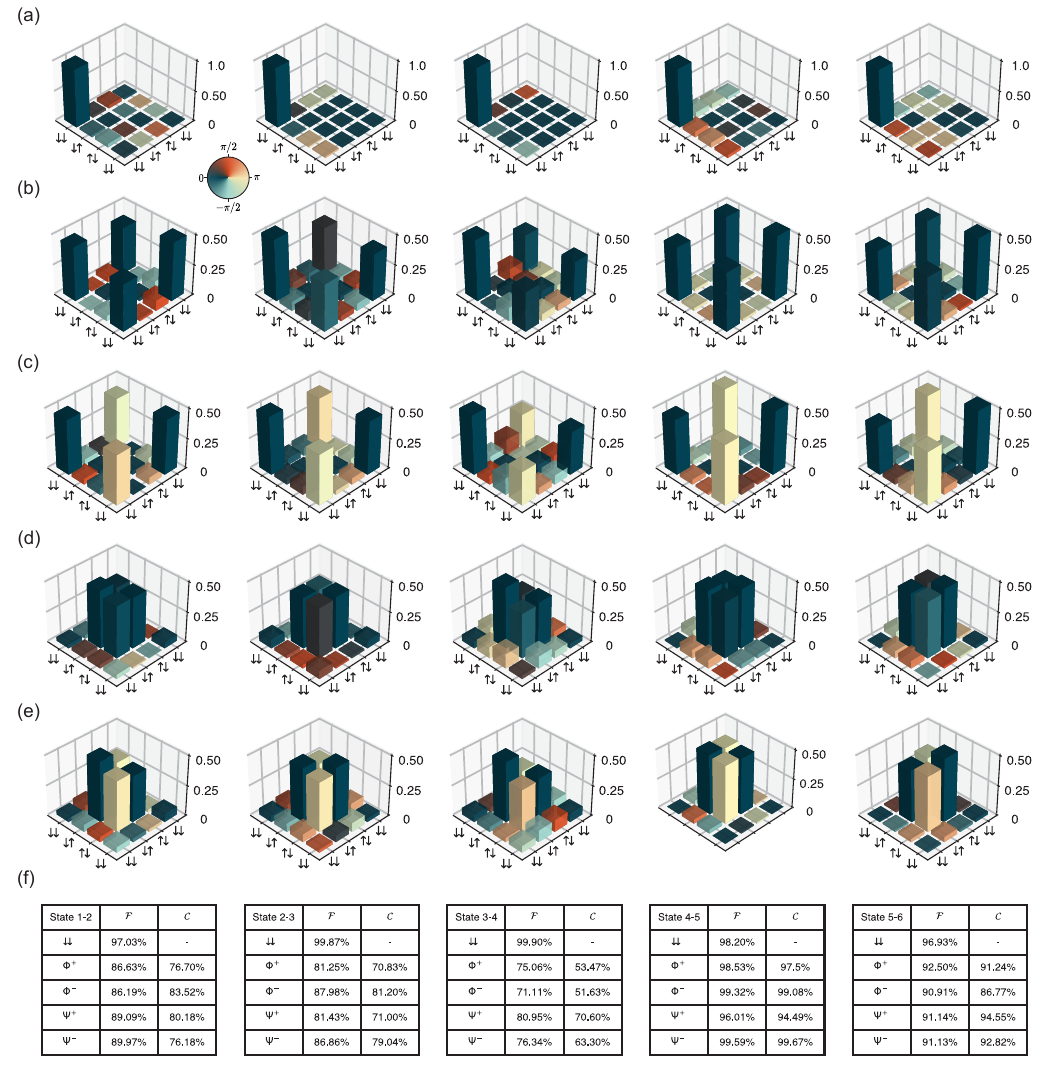}
  \caption{\textbf{Quantum state tomography of two-qubit states}. We prepare (\textbf{a}) the $\ket{\downarrow\downarrow}$ state and the four Bell states (\textbf{b}) $\Phi^+$, (\textbf{c}) $\Phi^-$, (\textbf{d}) $\Psi^+$, (\textbf{e}) $\Psi^-$ for each qubit pair along the array. State fidelities $\mathcal{F}$ and concurrences C with SPAM removal are displayed in the tables below the figures  following the colum order. }
  \label{bells}
\end{figure*}
\begin{figure*}[h!]
  \centering
  \includegraphics[width=\textwidth]{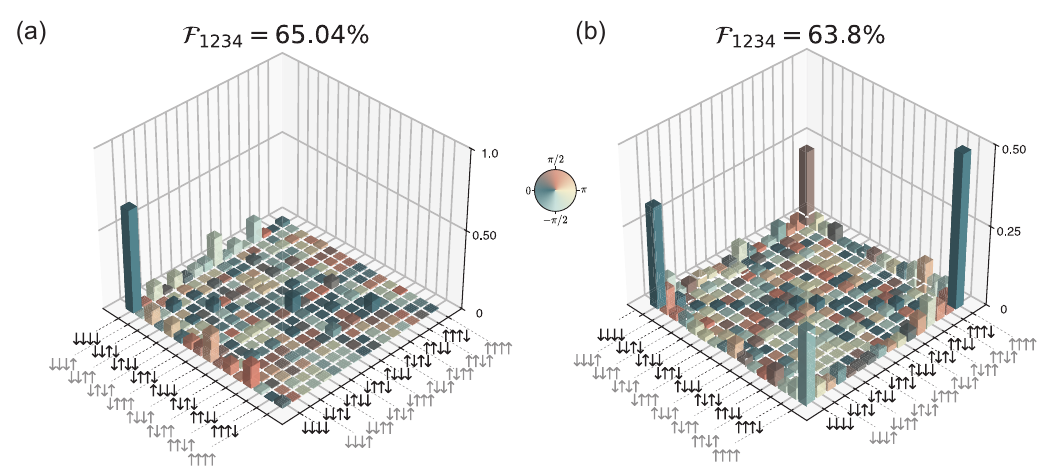}
  \caption{\textbf{Quantum state tomography of four-qubit states}. We prepare (\textbf{a}) the initial $\ket{\downarrow\downarrow\downarrow\downarrow}$ state , and (\textbf{b}) a GHZ state for qubits 1-2-3-4. State fidelities are displayed on top of the figures, which are obtained after removing SPAM errors.The fidelity of the GHZ state is obtained by fitting the reconstructed state to $\psi = (\ket{\downarrow\downarrow\downarrow\downarrow} + e^{i\theta}\ket{\uparrow\uparrow\uparrow\uparrow})/\sqrt{2}$, where $\theta = 1.71\pi$ is an arbitrary phase chosen to maximize $\mathcal{F}_{1234}$.}
  \label{fig:fourqubitdenmat}
\end{figure*}

The full density matrix can then be reconstructed following the approach in \cite{philips2022universal}. Crucially, the native ZZ measurement operator of the outer qubit pairs is mapped to a ZI/IZ operator to measure the single qubit projections. SPAM removal was done in a similar way as in \cite{philips2022universal}, based on the visibility of single-qubit Rabi oscillations. As this method does not remove all SPAM errors, the actual state fidelity would probably be higher than the ones shown. Complementing the three-qubit quantum state tomography results shown in Fig.\,4, we provide also results for various two-qubit qubit states in Fig.~\ref{bells} and for four-qubit states in Fig.~\ref{fig:fourqubitdenmat}.

%